%% file: main.tex
\documentclass[proof]{pasj01}
\Received{$\langle$reception date$\rangle$}
\Accepted{$\langle$acception date$\rangle$}
\Published{$\langle$publication date$\rangle$}

\usepackage{longtable}
\usepackage{threeparttablex}
\usepackage{booktabs}        
\usepackage{natbib}

\begin{document}

\title{Systematic study for gas-to-dust ratio of short gamma-ray burst afterglows}
\author{Kazuki Yoshida$^{*, \dag}$, Daisuke Yonetoku$^{*}$, Makoto Arimoto, Tatsuya
Sawano, Yasuaki Kagawa$^{\dag}$}%
\altaffiltext{}{Faculty of Mathmatics and Physics, Kanazawa University,
Kakuma-machi, Kanazawa, Ishikawa 920-1192}
\email{yoshida@astro.s.kanazawa-u.ac.jp, yonetoku@astro.s.kanazawa-u.ac.jp}

\maketitle

 \begin{abstract}
  Extra-galactic X-ray absorption and optical extinction are often found
  in gamma-ray burst (GRB) afterglows and they could be tracers of
  both circumburst and host galaxy environments. By performing spectral
  analyses for spectral energy distribution of 9~short GRB (SGRB)
  afterglows with known redshift, we investigated a ratio of the
  equivalent hydrogen column density to the dust extinction, $N_{\rm
  H}^{\rm rest}/A_{\rm V}^{\rm rest}$, in the rest frame of each SGRB. We
  found that the distribution of $N_{\rm H}^{\rm rest}/A_{\rm V}^{\rm
  rest}$ is systematically smaller than the one for long GRBs, and is
  roughly consistent with the gas-to-dust ratio in the Milky Way.
  This result means that the measured gas-to-dust ratio of SGRBs would
  originate from the interstellar medium in each host galaxy. This
  scenario supports the prediction that SGRBs occur in non star-forming
  regions in the host galaxies.
  \end{abstract}

 \footnotetext[$^{\dag}$]{JSPS Research Fellow}

\section{Introduction}
Gamma-ray bursts (GRBs) are grouped in two classes based on their
observed duration and spectral hardness of prompt emissions. Long GRBs
(LGRBs) and short GRBs (SGRBs) typically have duration of longer and
shorter than about 2~s, and relatively softer and harder spectra,
respectively \citep[e.g.,][]{Kouveliotou1993,Lien2016}. LGRBs are
almost always found in star-forming regions within
star-forming galaxies \citep{Bloom2002,Fruchter2006,Sevensson2010}, and
their progenitors are confirmed as the death of massive stars
\citep[e.g.,][and references therein]{Hjorth2003,Woosley2006,Kumar2015}.
On the other hand, some fraction of SGRBs occur in elliptical galaxies
showing no star-formation \citep{Fong2013a,Fong2013b}. The progenitors
of SGRBs are considered to be the coalescence of binary neutron star
(NS) and/or black hole (BH)-NS binary
\citep[e.g.,][]{Eichler1989,Narayan1992}.
In fact, the binary NS merger
event, GW~170817, was observed through the gravitational waves by the
LIGO and Virgo collaboration, which accompanied the SGRB candidate,
GRB~170817A, \citep{Abbott2017a,Goldstein2017,Savchenko2017}.
Since the binary system should move away from their birth cite until
its merging by natal kicks in the compact binary merger scenario
\citep[e.g.,][]{Narayan1992,Bloom1999,Fryer1999,Belczynski2006}, SGRBs
may occur in non star-forming regions inside of host galaxies or outside
of that. Therefore to investigate the
surrounding environment of SGRBs and compare it with that of LGRBs are
crucial way to interpret the SGRBs' progenitors.

To study the spectral energy distributions (SEDs) of GRB afterglows is
the major approach to interpret surrounding environments of GRBs. GRB
afterglows are thought to originate from relativistically expanding jets
that form shocks between the jet and the surrounding medium
\citep[e.g.,][]{Rees1992, Rees1998}, and their SEDs in the optical to
X-ray band can be described by a single or broken power-law function
\citep{Sari1998,Granot2002}. Performing the spectral analysis for them,
we can study extinction curves following SEDs and measure the
amounts of X-ray absorption and optical extinction in the host galaxy,
which are usually defined as an equivalent hydrogen column density
($N_{\rm H}$) under the assumption of the solar abundance and an extinction
in V band ($A_{\rm V}$), respectively. The extinction curve shows the
dependence of dust attenuation on wavelength, which originates from the dust
size and their chemical properties and are different for galaxies,
e.g., the Milky Way (MW), the Large Magellanic Cloud (LMC) and the Small
Magellanic Cloud (SMC), \citep[e.g.,][]{Pei1992}. The $N_{\rm H}/A_{\rm
V}$ ratios, called gas-to-dust ratio\footnote{This is sometimes
called metal-to-dust ratio, especially when the equivalent hydrogen
column density is derived from the X-ray absorption, because the
dominant X-ray absorbers are strictly metallic elements.}, reflect the
properties of the interstellar medium (ISM) in the galaxies and is
considered to vary with galaxies, e.g. the MW, LMC and SMC \citep{Welty2012}.

According to previous studies for afterglows of LGRBs
\citep[e.g.,][]{Schady2007, Schady2010, Covino2013}, in the optical and
near infrared (NIR) band, the extinction curve of the SMC well fits to
SEDs of observation data rather than the one of the MW or LMC in almost all
events. However, in the rest frame of each GRB, the ratio of hydrogen
equivalent column density measured in X-ray band to the
dust extinction measured in optical/NIR band ($N_{\rm H}^{\rm
rest}/A_{\rm V}^{\rm rest}$) is significantly larger than the ones in
the SMC as well as the MW and LMC. The dust destruction caused by
the intense GRB emission is discussed as an major interpretation of the large
$N_{\rm H}^{\rm rest}/A_{\rm V}^{\rm rest}$, but its observational
evidence has been not found
\citep{Waxman2000,Galama2001,Savaglio2003,Schady2010}. \citet{Schady2010}
reports the possibility that the $N_{\rm H}/A_{\rm V}$ ratio of LGRBs in
low-metallicity galaxies is large. On the other hand, \citet{Zafar2011}
investigated the $N_{\rm H}^{\rm rest}/A_{\rm V}^{\rm rest}$ ratio
including metallicity of each LGRB in detail, but they concluded that
only the metallicity can not explain the observed high $N_{\rm H}^{\rm
rest}/A_{\rm V}^{\rm rest}$ ratio. Until now, a unified picture to
explain such a large $N_{\rm H}^{\rm rest}/A_{\rm V}^{\rm rest}$ ratio
has not been established.

In this paper, we systematically performed SED fitting for 9~SGRBs with
known redshifts using both X-ray and optical/NIR afterglow data, and
investigated the ratio of equivalent hydrogen column density to optical
extinction of each GRB. Furthermore, we compared these ratio with the
results of LGRBs and also typical galaxy environment. The error and
upper/lower limits of all fitting parameters are shown at $68\%$ and
$90\%$ confidence level, respectively.

\section{Data reduction and analysis}
We used SGRBs with known redshifts observed by the X-ray Telescope (XRT)
on board the {\it Neil Gehrels Swift Observatory} ({\it Swift})
\citep{Gehrels2004,Burrows2005}. In addition to obvious SGRBs with $T_{90}<2$~sec,
we included possible SGRB candidates with $T_{90}>2$~sec, which are
considered as the SGRB with extended soft X-ray emissions following 
prompt emissions. Here $T_{90}$ is the time duration which includes 90\% of the
observed photon counts except for the first and the last 5\% in the GRB
emission observed the {\it Swift}/BAT. We selected
brighter 9~SGRBs, listed in Table \ref{tab:GRBlist}, whose host galaxies
were much dimmer than the optical/NIR afterglows.

Since the spectral parameters of the power-law index and
the dust extinction in the SEDs of GRB afterglows are degenerate, we
cannot correctly measure the dust extinction in the rest frame of SGRBs
with only optical/NIR data, which are limited data points. Therefore, in
order to obtain the reliable spectral parameters, we performed the
simultaneous spectral analysis for broadband SEDs consist of both
optical/NIR and X-ray data, i.e. we estimate the spectral index in
optical/NIR band including X-ray data. In \citet{Covino2013}, the
optical extinctions derived from only optical/NIR data analysis were
consistent with those derived from the X-ray prior analysis as we
mentioned.

\subsection{Optical/NIR data}
We gathered available data (not including upper limits) of optical/NIR
afterglow observations from the published papers and GCN
Circulars\footnote{https://gcn.gsfc.nasa.gov/}, and converted their
magnitude to the flux density. The data we used and the references of
them are listed in Table \ref{tab:opt}. Using the database in the
NASA/IPAC Infrared Science
Archive\footnote{https://irsa.ipac.caltech.edu/applications/DUST/}
\citep{Schlafly2011}, we converted the observed flux density of each
burst to the one before affecting the galactic extinction.

Since the GRB afterglow shows power-law decline in time
\citep{Sari1998,Granot2002}, it is necessary to collect data at the same
time as close as possible in order to create accurate
SED. Here, we ignore the time difference among each band
data observed almost at the same time (or slightly different time) when
the relative uncertainty of the measured flux density ($\Delta F/F$) and
the observation time ($\Delta t/t$) satisfies $\Delta F/F > \Delta
t/t$. Since six of the nine samples satisfied the condition, we used the
observation data of that epoch as the SEDs for these events. For the
other three samples, GRB~070724A, 090510 and 140903A, we adopted a
power-law function of $F(t) \propto (t - t_0)^{\alpha_{\rm opt}}$ to the
observed light curve in the same band, and we estimate the flux density
at the time when the interpolation and extrapolation in all bands are
minimized. Here, $t_0$ is the trigger time and $\alpha_{\rm opt}$ is the
temporal index in optical/NIR band. The time we set for each sample is
summarized in Table \ref{tab:GRBlist}.

\subsection{X-ray data}
X-ray observation data of SGRBs are taken from UK {\it Swift}
Science Data Centre\footnote{http://www.swift.ac.uk/index.php}.
The XRT observation is generally performed
in two modes, windowed timing (WT) and photon counting (PC) mode. The PC
mode data as data of afterglows were used in this analysis, since the
extended emission is often observed in the WT mode, whose origin is
different from the one of afterglows
\citep[e.g.,][]{Norris2006,Kagawa2015,Kisaka2017}. The light-curve data
were taken from the XRT light curve repository\footnote{http://www.swift.ac.uk/xrt\_curves/}
\citep{Evans2007,Evans2009}. We extracted a source
and background event data from circle region with 20 pixels and 40 in
radius (corresponding to 47 and 94 arcsec), respectively, which are
recommended ones in the {\it Swift} XRT Users Guide Version
1.2\footnote{https://swift.gsfc.nasa.gov/analysis/}. Using {\tt XSELECT}
software
(v2.4)\footnote{https://heasarc.gsfc.nasa.gov/docs/software/lheasoft/ftools/xselect/},
we extracted spectral data from the cleaned event data. For the spectral
analysis, ancillary response files were created by {\tt xrtmkarf}
(v0.6.3) and response matrices were taken from the calibration database
files\footnote{https://heasarc.gsfc.nasa.gov/docs/heasarc/caldb/swift/}.

In \citet{Kagawa2019}, they analyzed time-resolved X-ray spectra
whose time intervals were divided to each spectrum
contains 128 photons, and
spectral parameters at each time were obtained. They also analyzed the
time averaged spectra with all observation data in PC mode, and
confirmed that the photon indices of both results are consistent with
each other within the error. Thus we performed time averaged spectral
analyses with the entire PC mode data to maximize a signal-to-noise
ratio. The time averaged spectra were grouped to 20 counts per energy
bin.

\begin{figure*}
 \begin{center}
  \begin{tabular}{c}
   \begin{minipage}{0.333\hsize}
    \begin{center}
     \includegraphics[width=\hsize,clip,viewport=10 10 535 528]{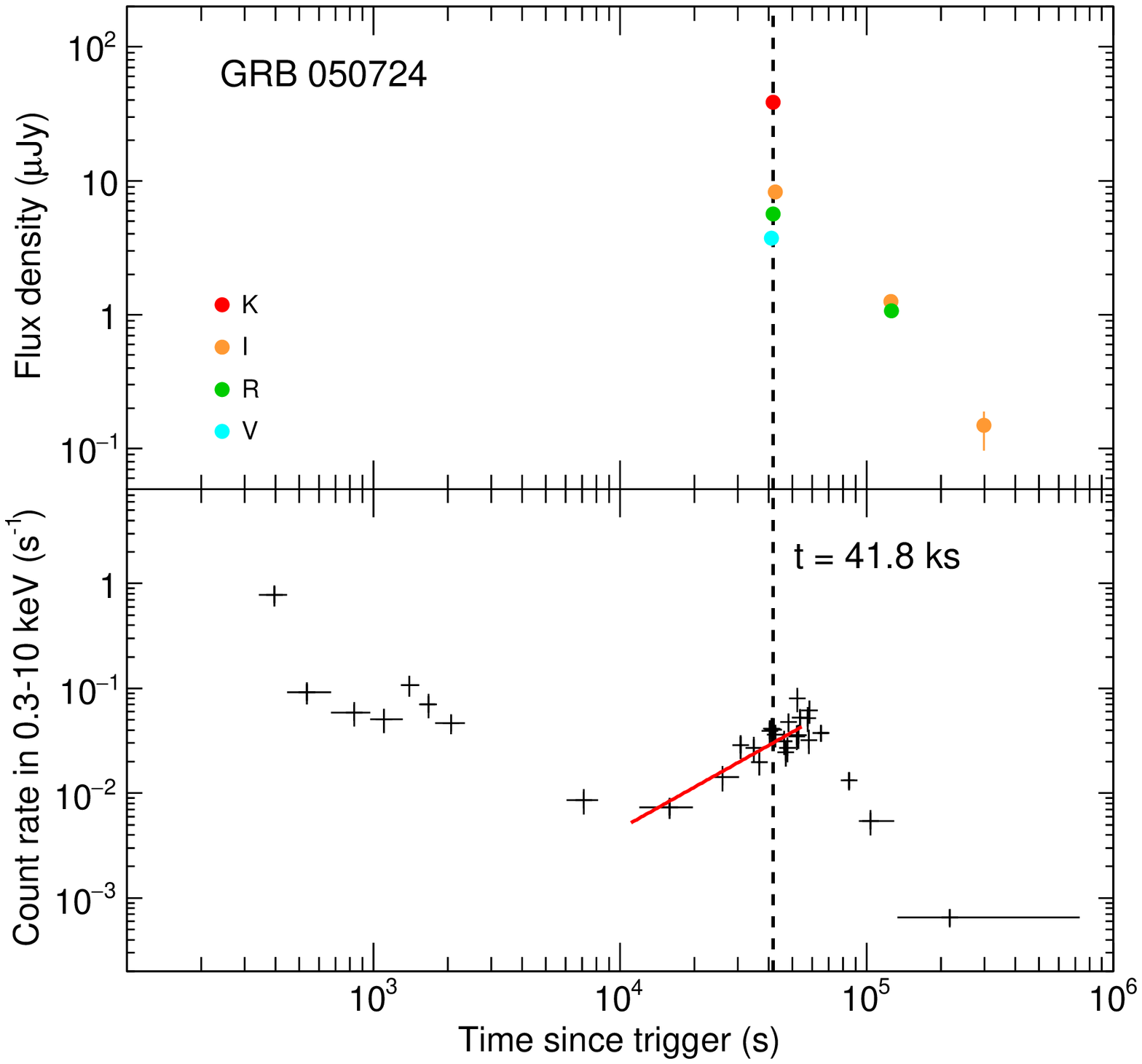}
    \end{center}
   \end{minipage}
   \begin{minipage}{0.333\hsize}
    \begin{center}
     \includegraphics[width=\hsize,clip,viewport=10 10 535 528]{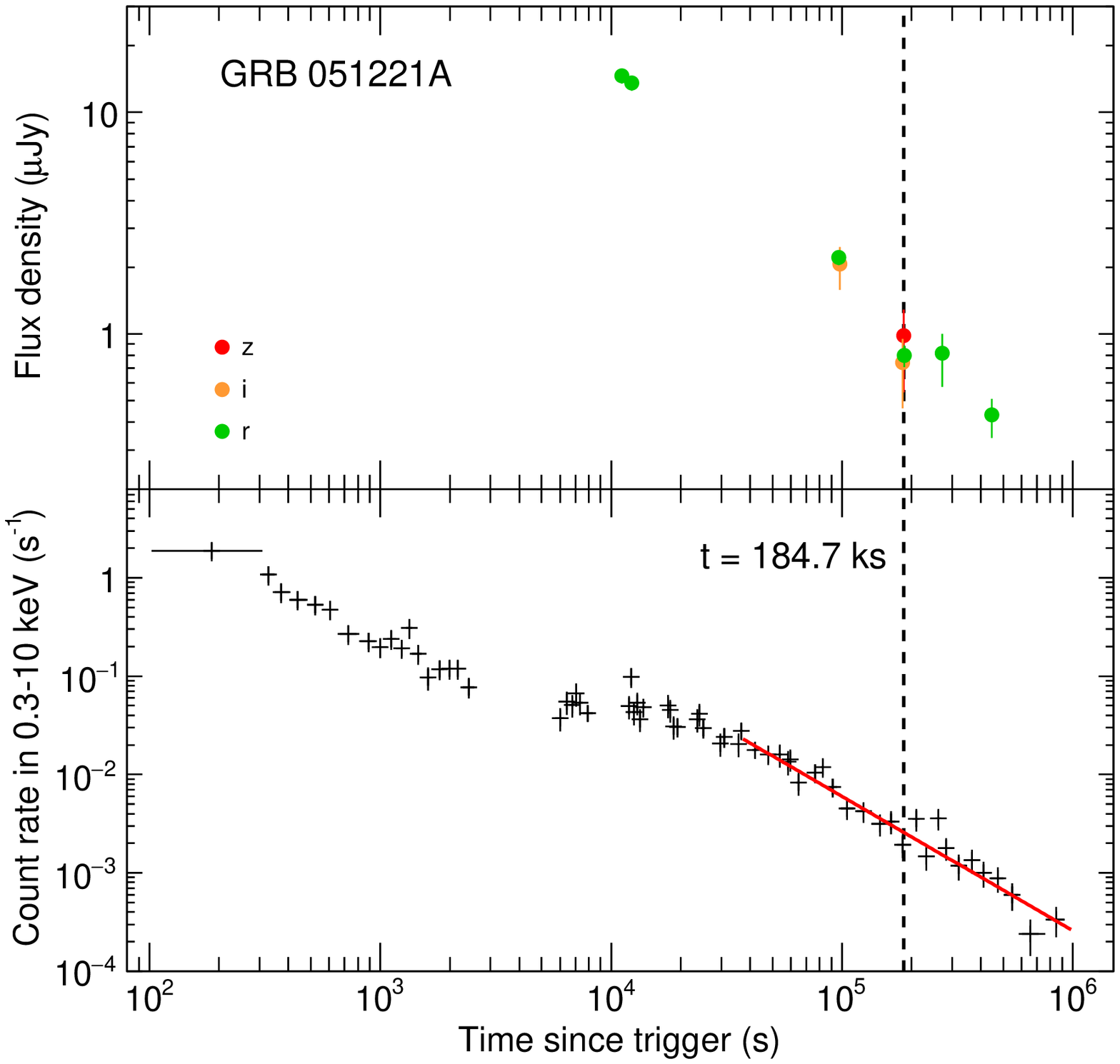}
    \end{center}
   \end{minipage}
   \begin{minipage}{0.333\hsize}
    \begin{center}
     \includegraphics[width=\hsize,clip,viewport=10 10 535 528]{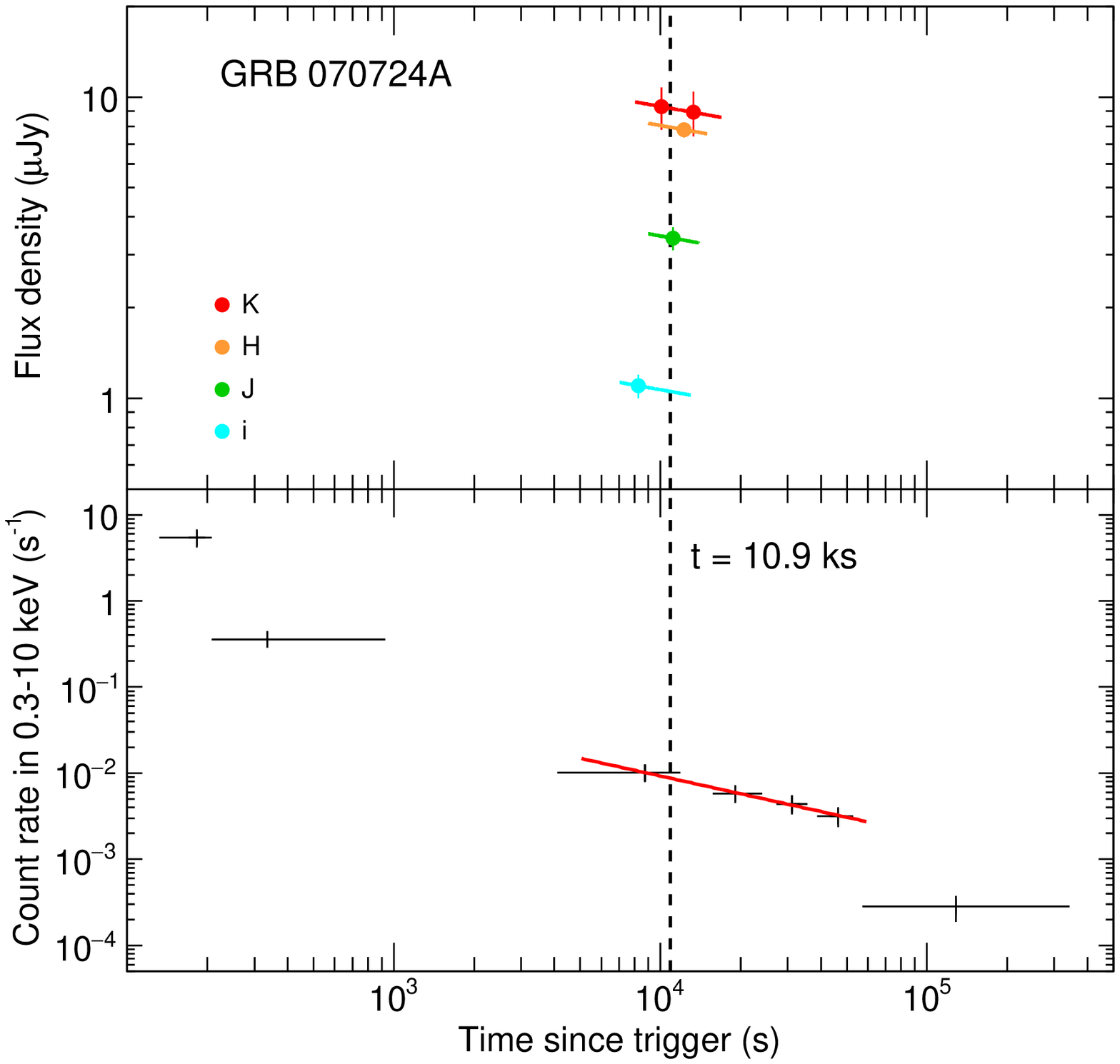}
    \end{center}
   \end{minipage}\\
   \begin{minipage}{0.333\hsize}
    \begin{center}
     \includegraphics[width=\hsize,clip,viewport=10 10 535 528]{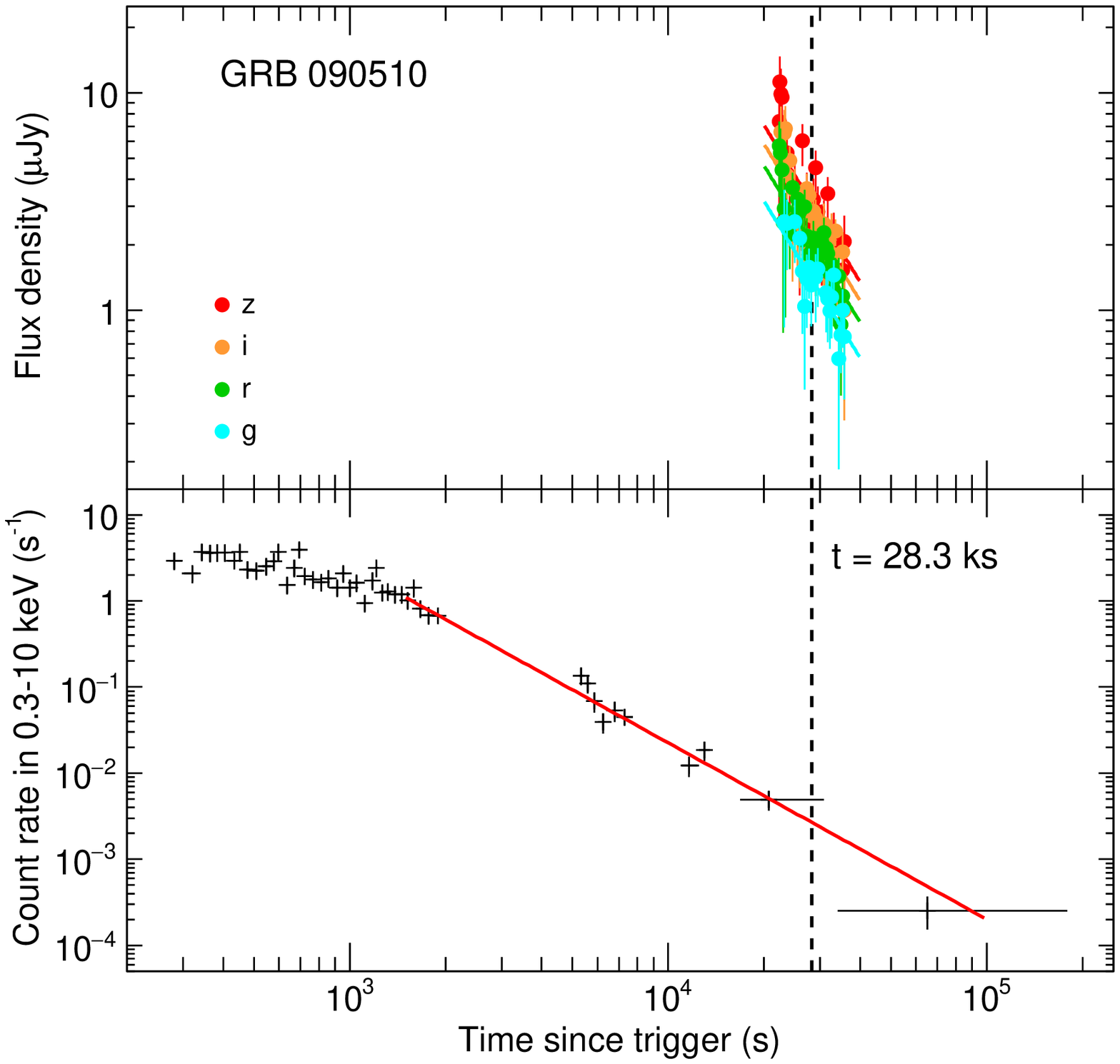}
    \end{center}
   \end{minipage}
   \begin{minipage}{0.333\hsize}
    \begin{center}
     \includegraphics[width=\hsize,clip,viewport=10 10 535 528]{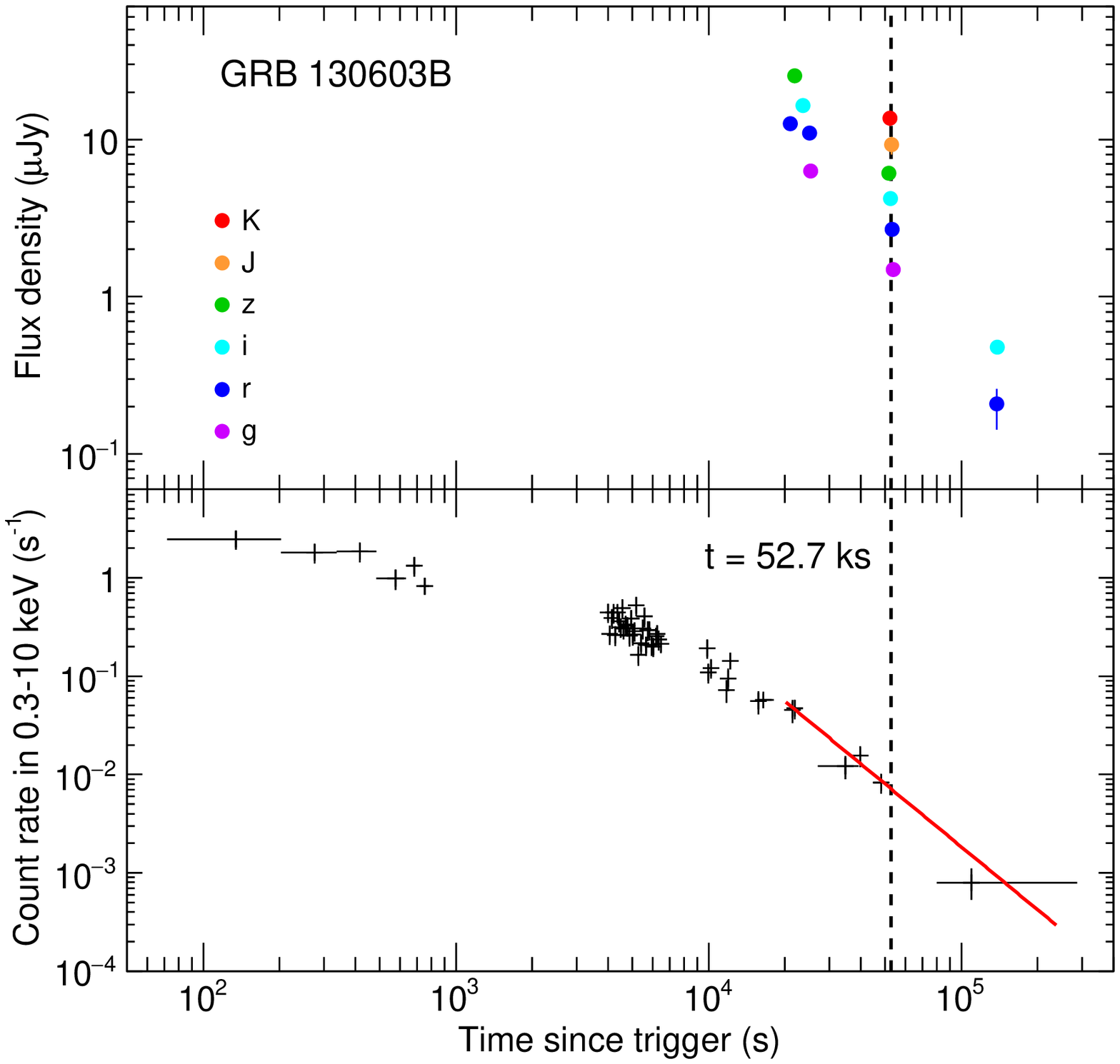}
    \end{center}
   \end{minipage}
   \begin{minipage}{0.333\hsize}
    \begin{center}
     \includegraphics[width=\hsize,clip,viewport=10 10 535 528]{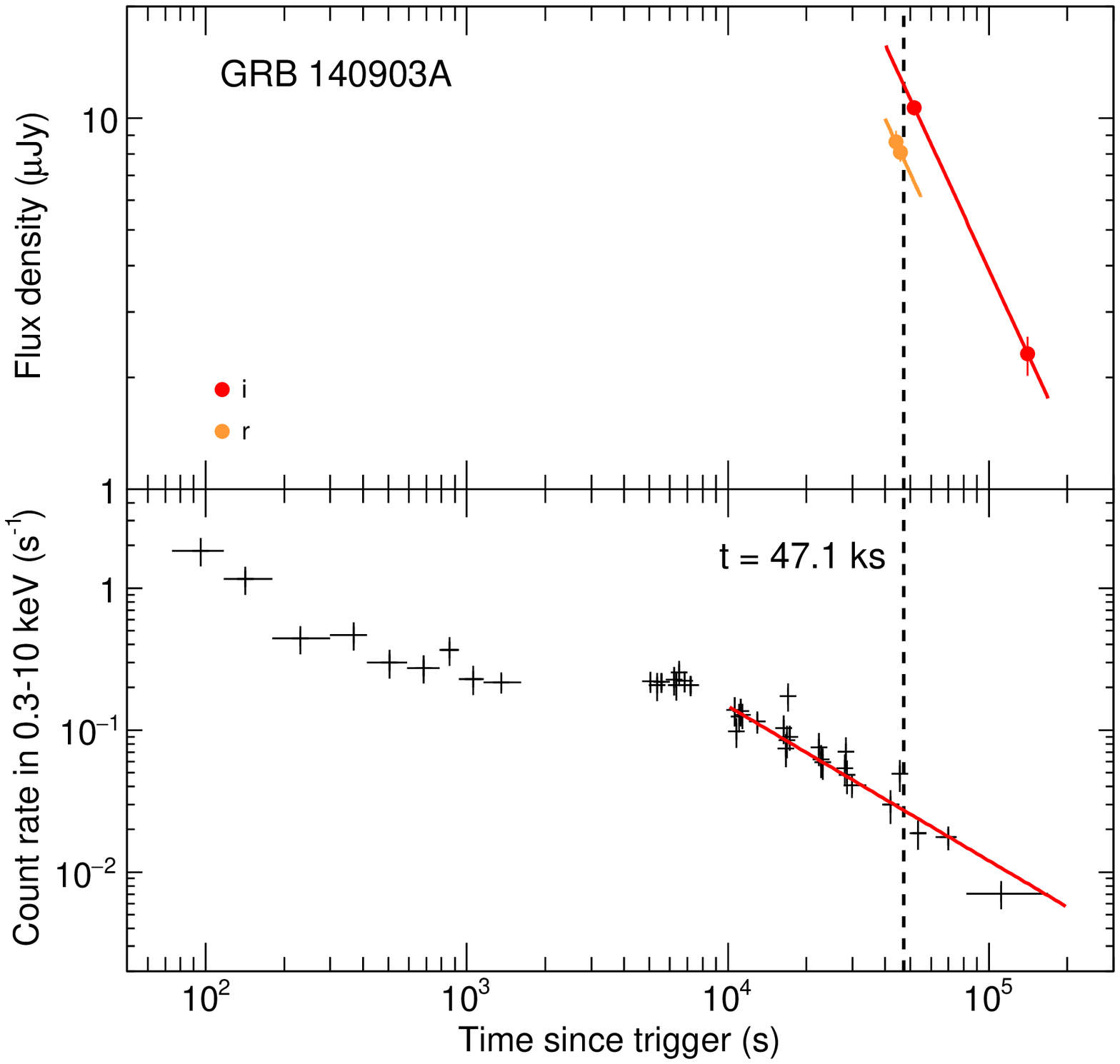}
    \end{center}
   \end{minipage}\\
   \begin{minipage}{0.333\hsize}
    \begin{center}
     \includegraphics[width=\hsize,clip,viewport=10 10 535 528]{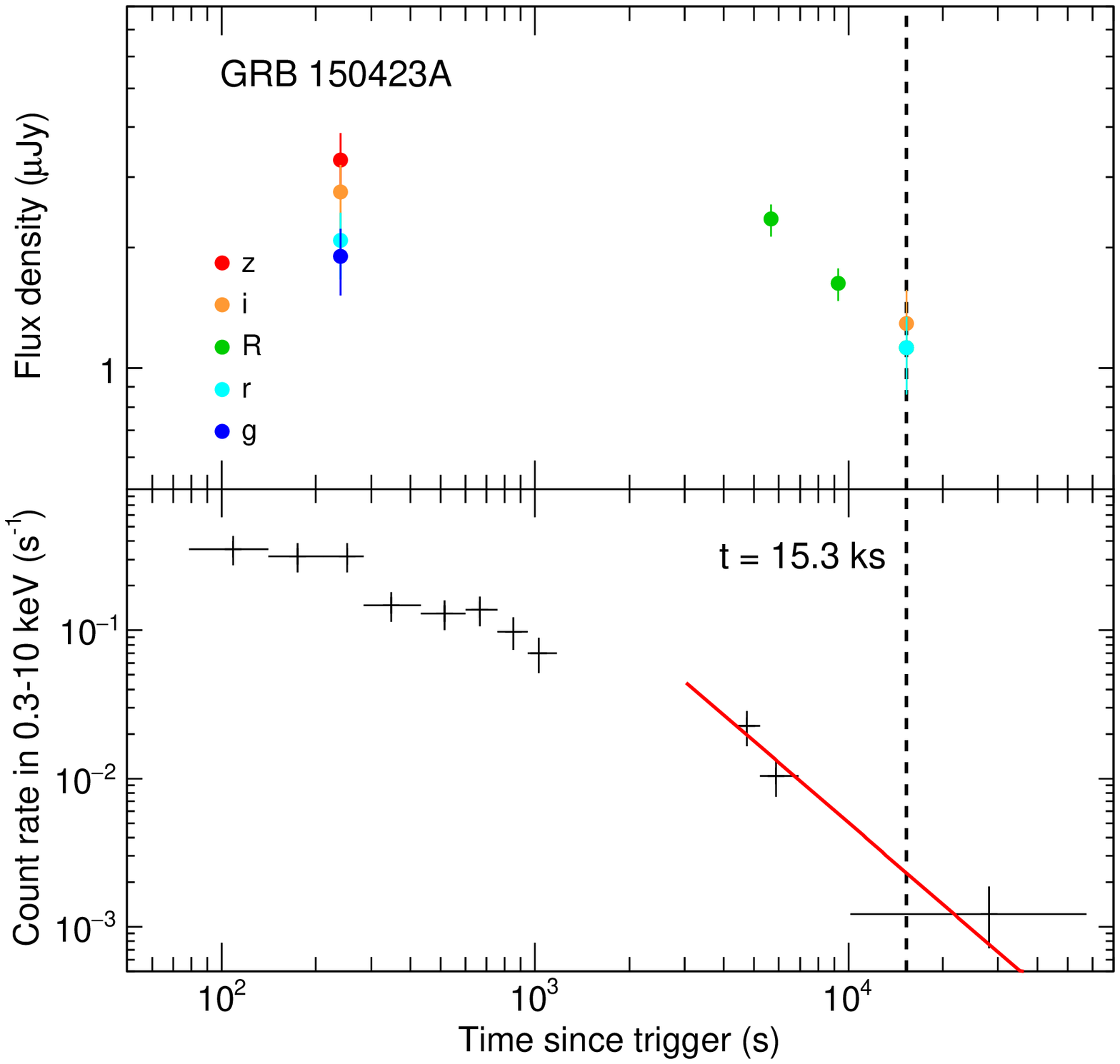}
    \end{center}
   \end{minipage}
   \begin{minipage}{0.333\hsize}
    \begin{center}
     \includegraphics[width=\hsize,clip,viewport=10 10 535 528]{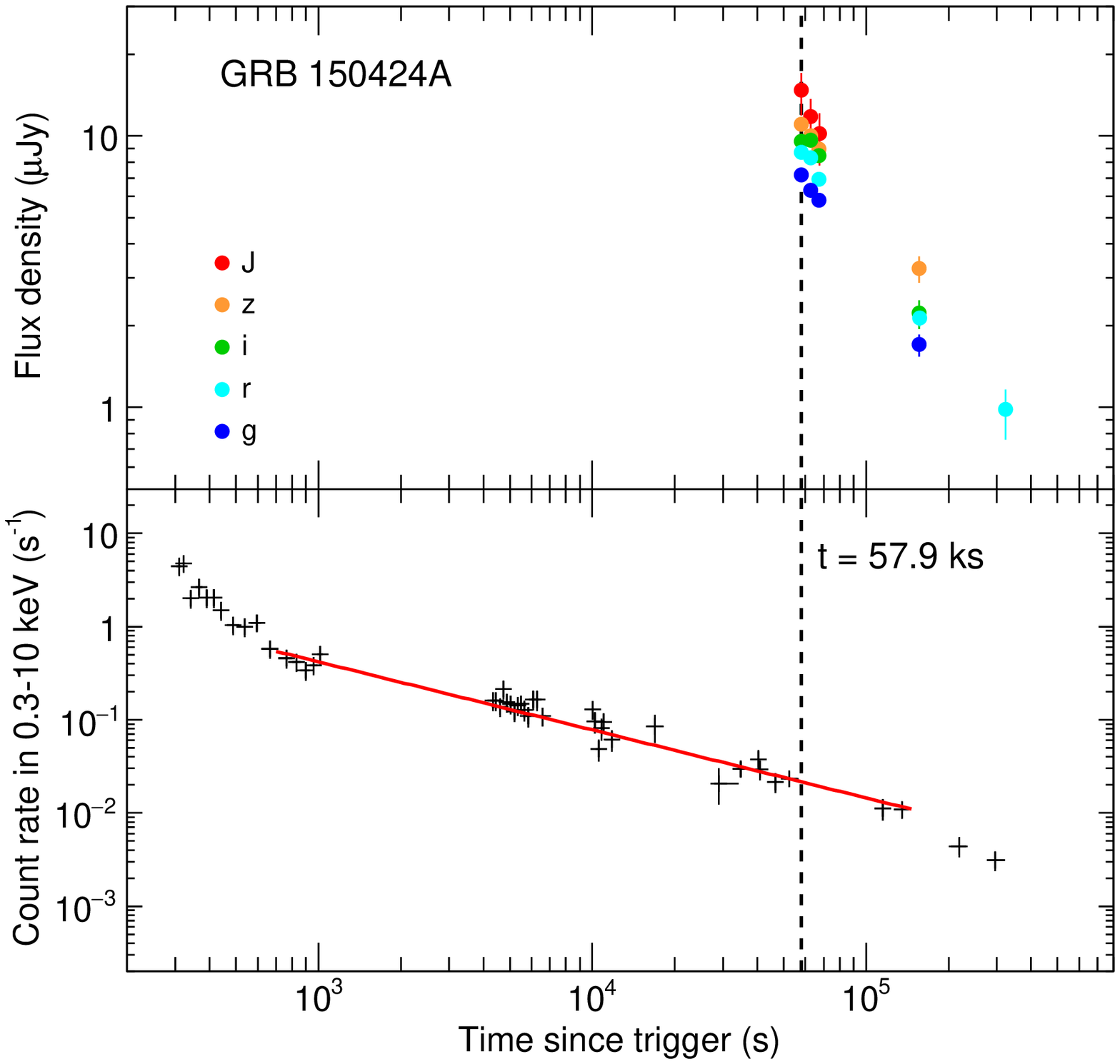}
    \end{center}
   \end{minipage}
   \begin{minipage}{0.333\hsize}
    \begin{center}
     \includegraphics[width=\hsize,clip,viewport=10 10 535 528]{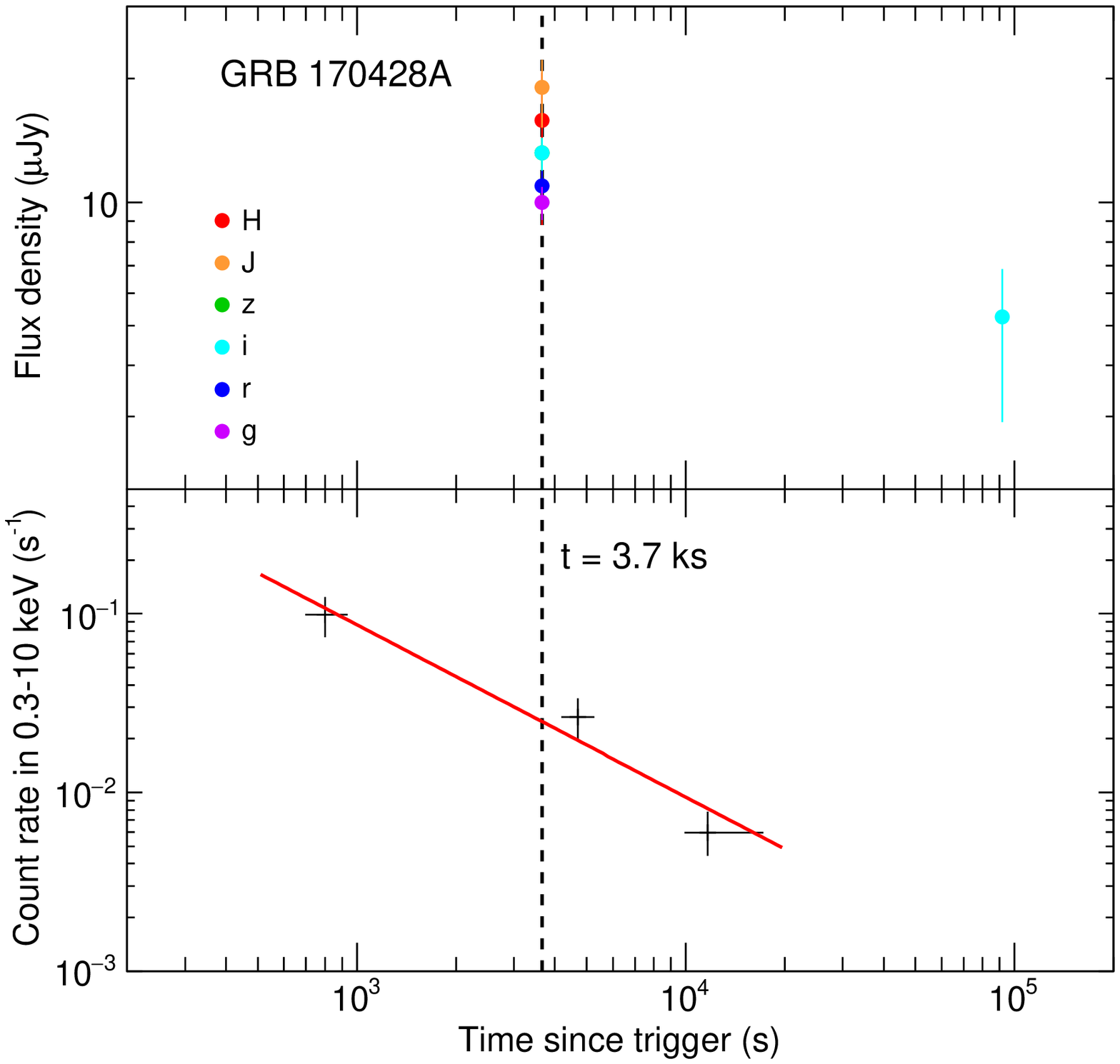}
    \end{center}
   \end{minipage}
  \end{tabular}
  \vspace{0.05in}
  \caption{Optical/NIR and X-ray light curves in the observer frame. The
  solid lines and the vertical dashed lines show the best-fit power-law
  models of each observation band and the epoch of broadband SEDs of
  each SGRB, respectively.\label{fig:lc}}
 \end{center}
\end{figure*}

In order to determine the X-ray flux at any given time, we adopted
the power-law function with the temporal index of X-ray band
($\alpha_{\rm X}$) to the X-ray light curves in the same way we did for
optical/NIR light curves. Where light curve data
were taken from the {\it Swift}-XRT lightcurve
repository\footnote{http://www.swift.ac.uk/xrt\_curves/}, in which the
systematic search of temporal breaks had been performed for light
curves \citep{Evans2007,Evans2009}. Considering their results and
excluding the time at the temporal breaks, we defined fitting intervals
with simple power-laws. The fitting results are shown in Figure
\ref{fig:lc} as red solid lines. Using the best-fitting result, we
estimated a conversion factor from average flux to the one of focusing
time and renormalized the time-averaged X-ray spectra for the broadband
SED analysis.

\subsection{Spectral analysis}
The spectral analysis is carried out with {\tt XSPEC}
software (v12.9.0)\footnote{https://heasarc.gsfc.nasa.gov/xanadu/xspec/}
and fit models prepared in there.
Based on a standard synchrotron shock model \citep{Sari1998, Granot2002},
we adopted a {\tt powerlaw} model and {\tt bknpower} model for the
broadband SEDs. The X-ray spectral index ($\beta_{\rm X}$) is derived
from the photon index ($\Gamma$) of power-law in the relation of $\beta_{\rm X} = 1
- \Gamma$. Then we imposed the spectral index of the optical/NIR region,
$\beta_{\rm opt} = \beta_{\rm X}$ in the {\tt powerlaw} model and
$\beta_{\rm opt} = \beta_{\rm X} - 0.5$ in the {\tt bknpower} model. The
latter case corresponds to the condition where the cooling frequency of
the synchrotron emission locates between the optical/NIR and X-ray ranges
\citep{Sari1998, Granot2002}.

We added {\tt phabs} and {\tt zphabs} models corresponding to the
photo-electric absorption in our galaxy and host galaxy,
respectively. The parameter of the Galactic equivalent hydrogen column
density ($N_{\rm H}^{\rm gal}$) is fixed to be the amount calculated for
the sky coordinates of each SGRB by the database in the UK Swift Science
Data Center\footnote{http://www.swift.ac.uk/analysis/nhtot/index.php}
\citep{Willingale2013}, as shown in Table \ref{tab:GRBlist}. The
equivalent hydrogen column density in the host galaxy ($N_{\rm
H}^{\rm rest}$) was derived from the model fit where the solar
abundances were assumed. We note that the metallicity of the SGRB host
galaxies show a wide value, but on the average, it is about a solar
abundance \citep[][and references therein]{Berger2014}.

To compute the extinctions in the host galaxy, we used the {\tt zdust}
model that considered extinction for wavelength by dust grains as
described in \citet{Pei1992}. There are major three models of the
extinction curves in the MW, LMC and SMC environments. We adopted all
three extinction models and investigated the difference of extinction in
each model. All results of our spectral analysis are summarized in Table
\ref{tab:result}, but in the Section 3, we reported the results of using
the MW extinction model because there is little difference of the
amount of optical extinction among the three models. In fact, the
three extinction models are almost the same within the wavelength
range of the observation data in the rest frame of 9 SGRBs.

\section{Results}
Figure \ref{fig:lc} shows the optical/NIR and X-ray light curves and the
epoch of the broadband SED of each GRB. Although the time when the multi-band
observation was performed for GRB~050724 is in the X-ray flare phase, we set this epoch
for the broadband SED because it is reported in \citet{Berger2005}
\citep[see also][]{Malesani2007} that the optical/NIR and X-ray
emission might belong to the same component. In GRB~150423A, there are
two times with multi-band observation data, i.e an early epoch ($\sim240$
s) and a later one ($\sim15300$ s). Since the extended emission was
observed in the early epoch \citep{Kisaka2017,Kagawa2019}, we selected
the later epoch.

\begin{figure*}
 \begin{center}
  \begin{tabular}{c}
   \begin{minipage}{0.33\hsize}
    \begin{center}
     \includegraphics[width=1.06\hsize]{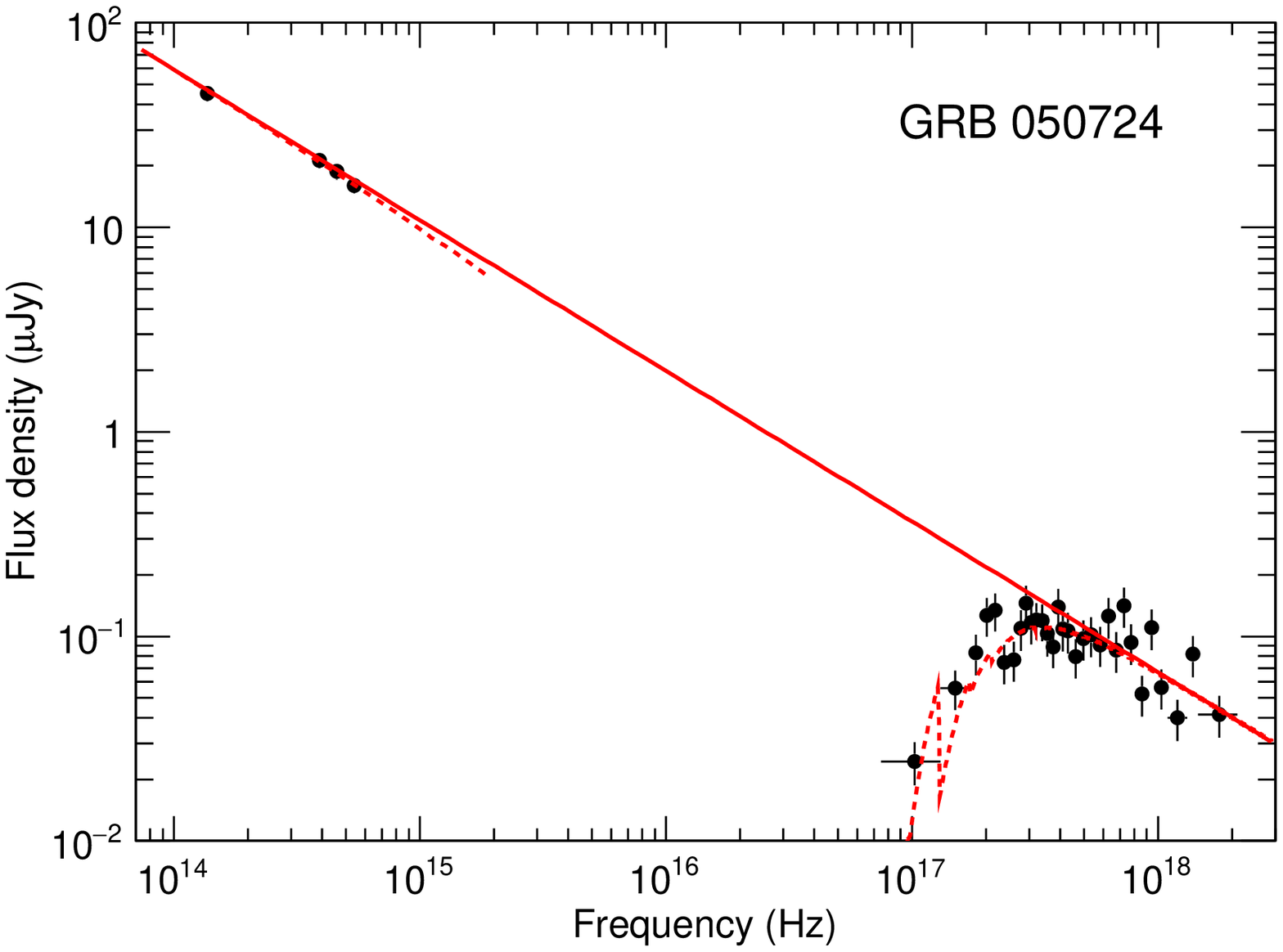}
    \end{center}
   \end{minipage}
   \begin{minipage}{0.33\hsize}
    \begin{center}
     \includegraphics[width=1.06\hsize]{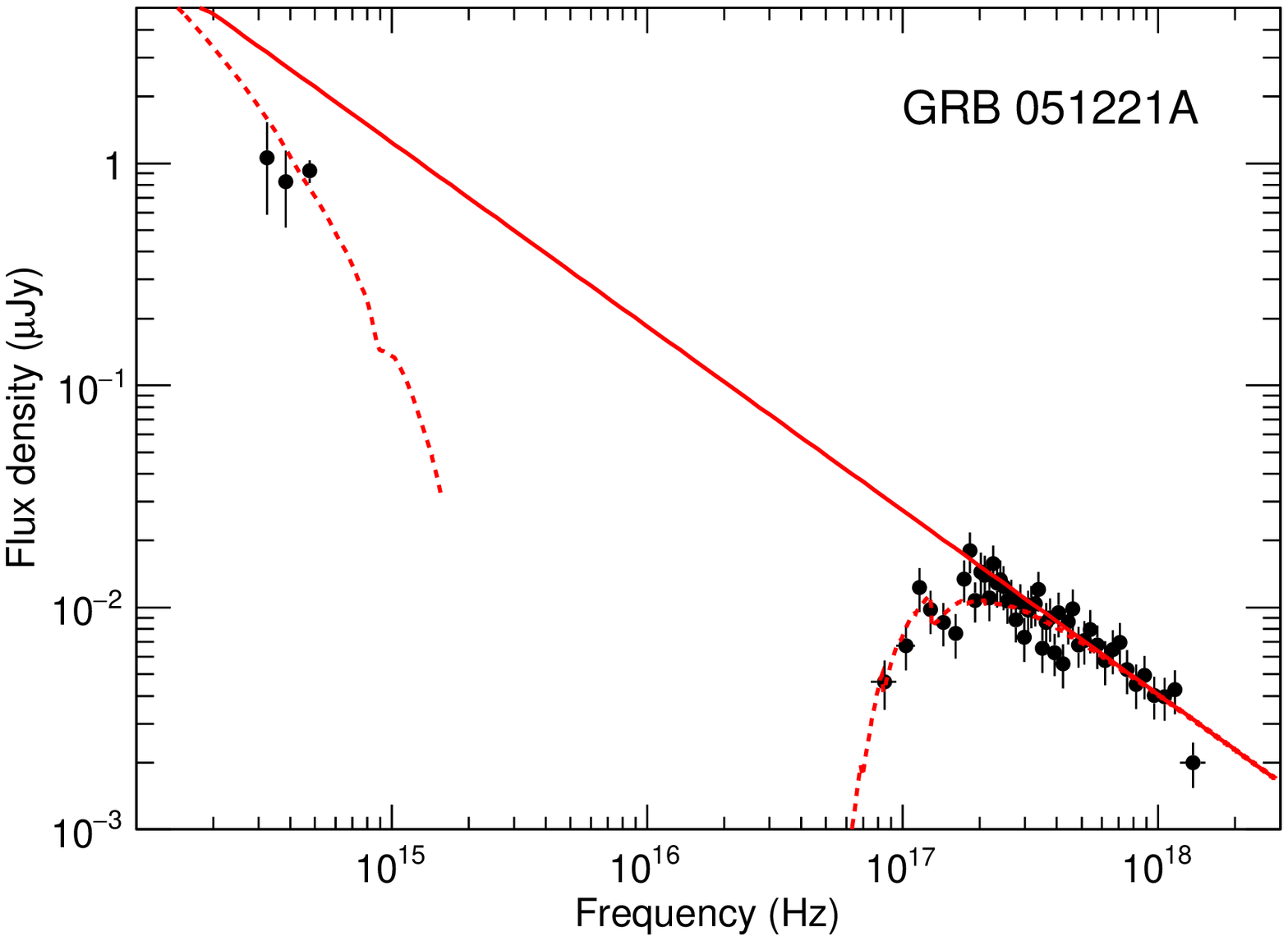}
    \end{center}
   \end{minipage}
   \begin{minipage}{0.33\hsize}
    \begin{center}
     \includegraphics[width=1.06\hsize]{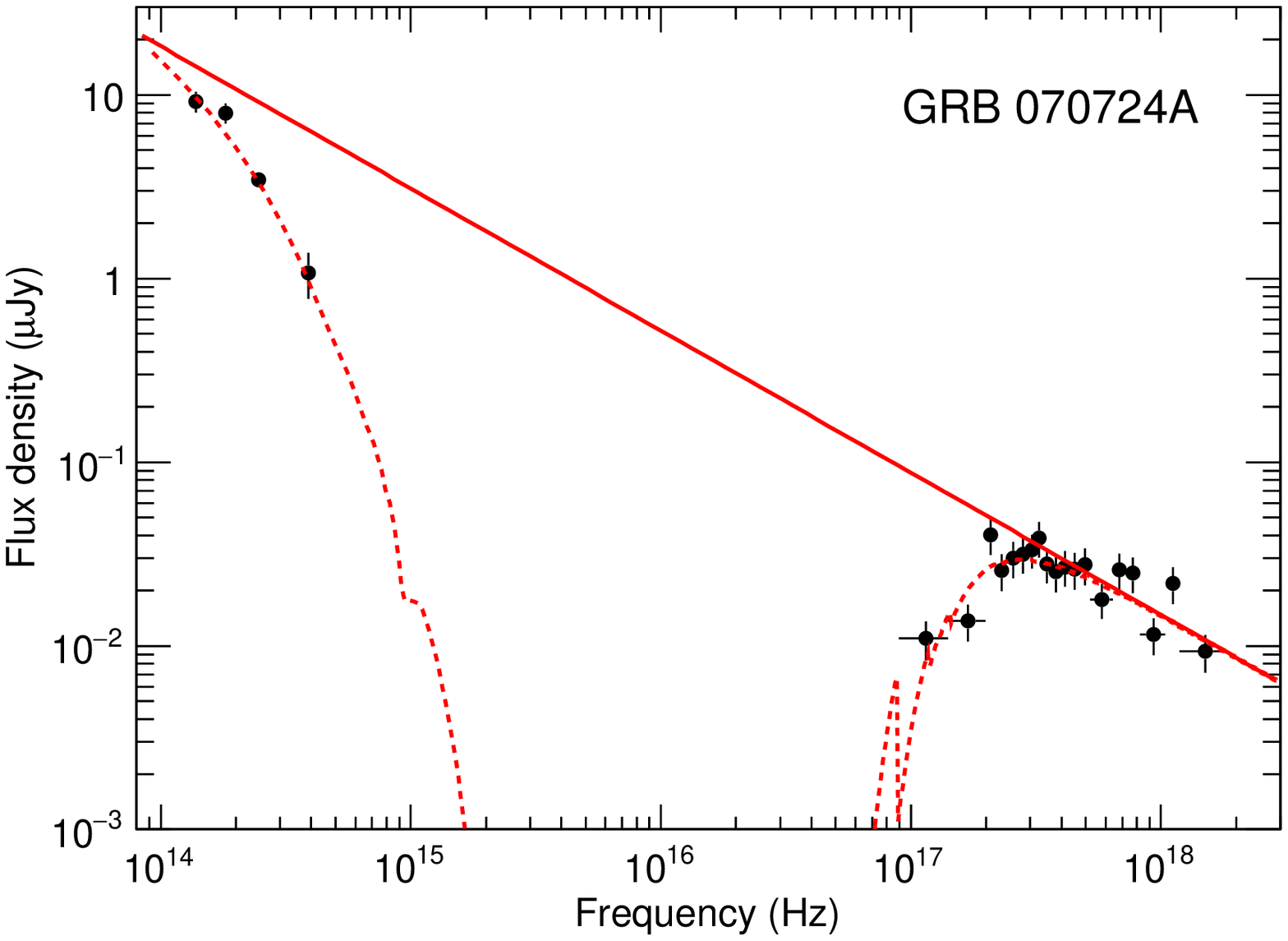}
    \end{center}
   \end{minipage}\\
   \begin{minipage}{0.33\hsize}
    \begin{center}
     \includegraphics[width=1.06\hsize]{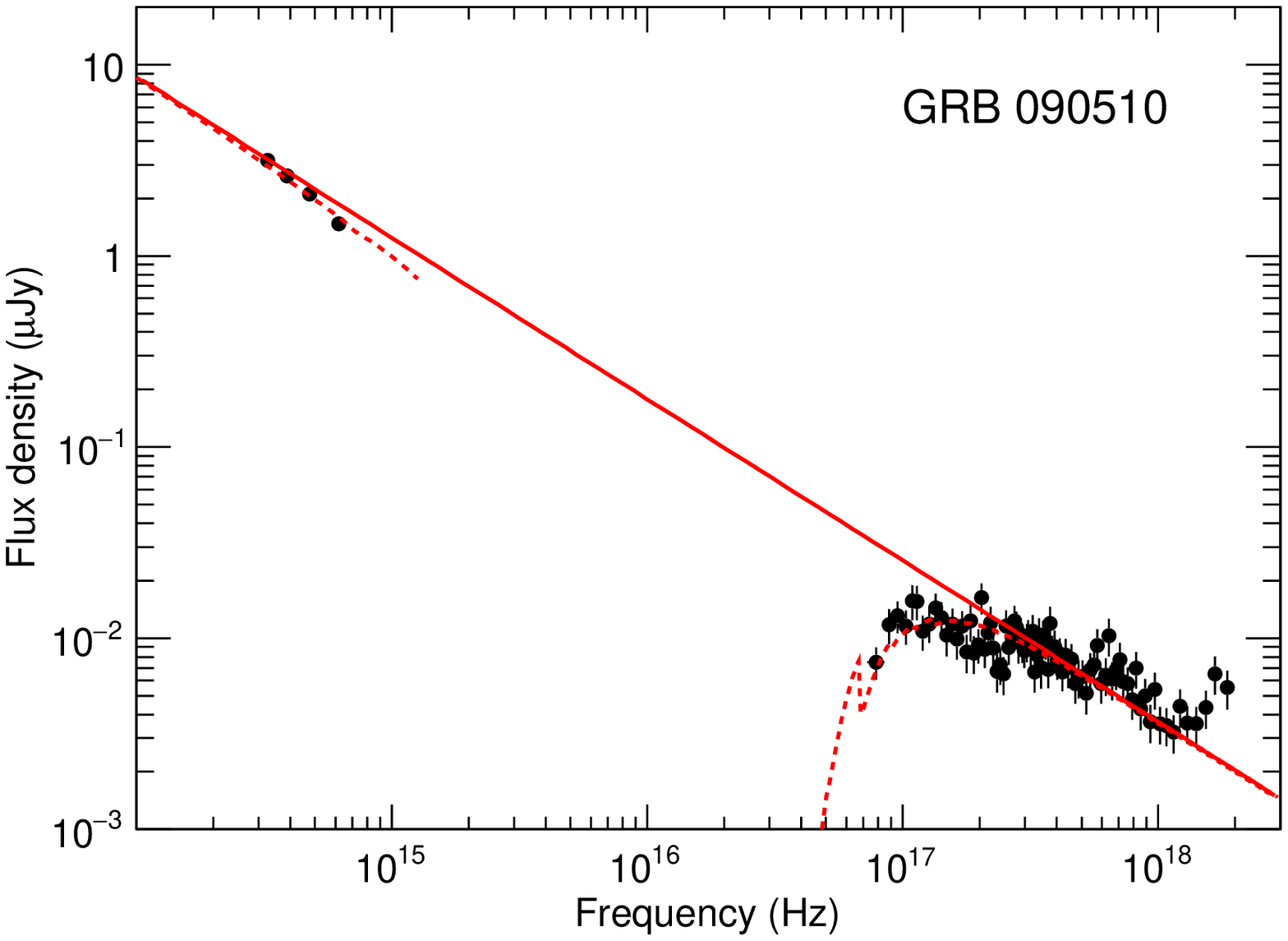}
    \end{center}
   \end{minipage}
   \begin{minipage}{0.33\hsize}
    \begin{center}
     \includegraphics[width=1.06\hsize]{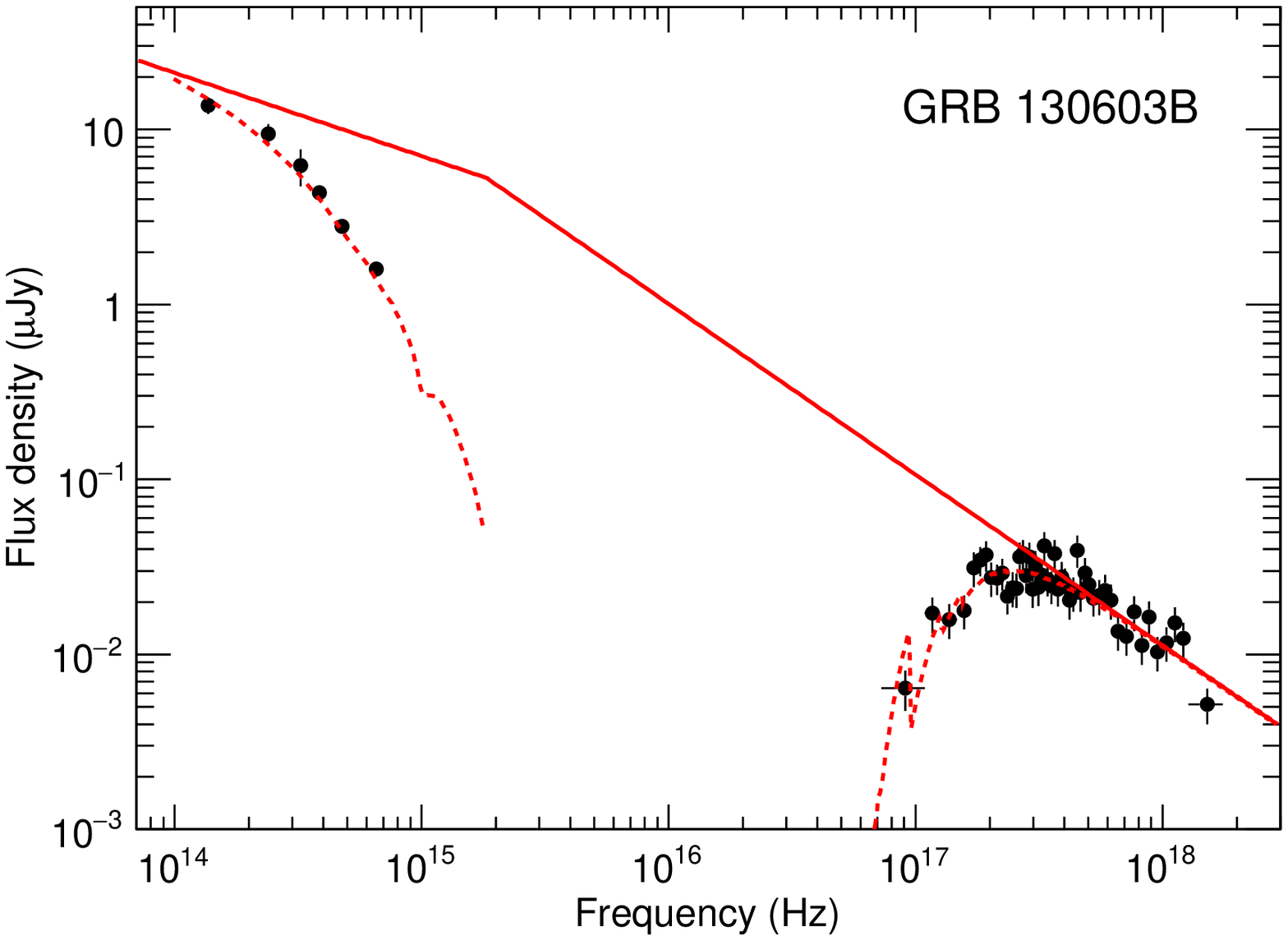}
    \end{center}
   \end{minipage}
   \begin{minipage}{0.33\hsize}
    \begin{center}
     \includegraphics[width=1.06\hsize]{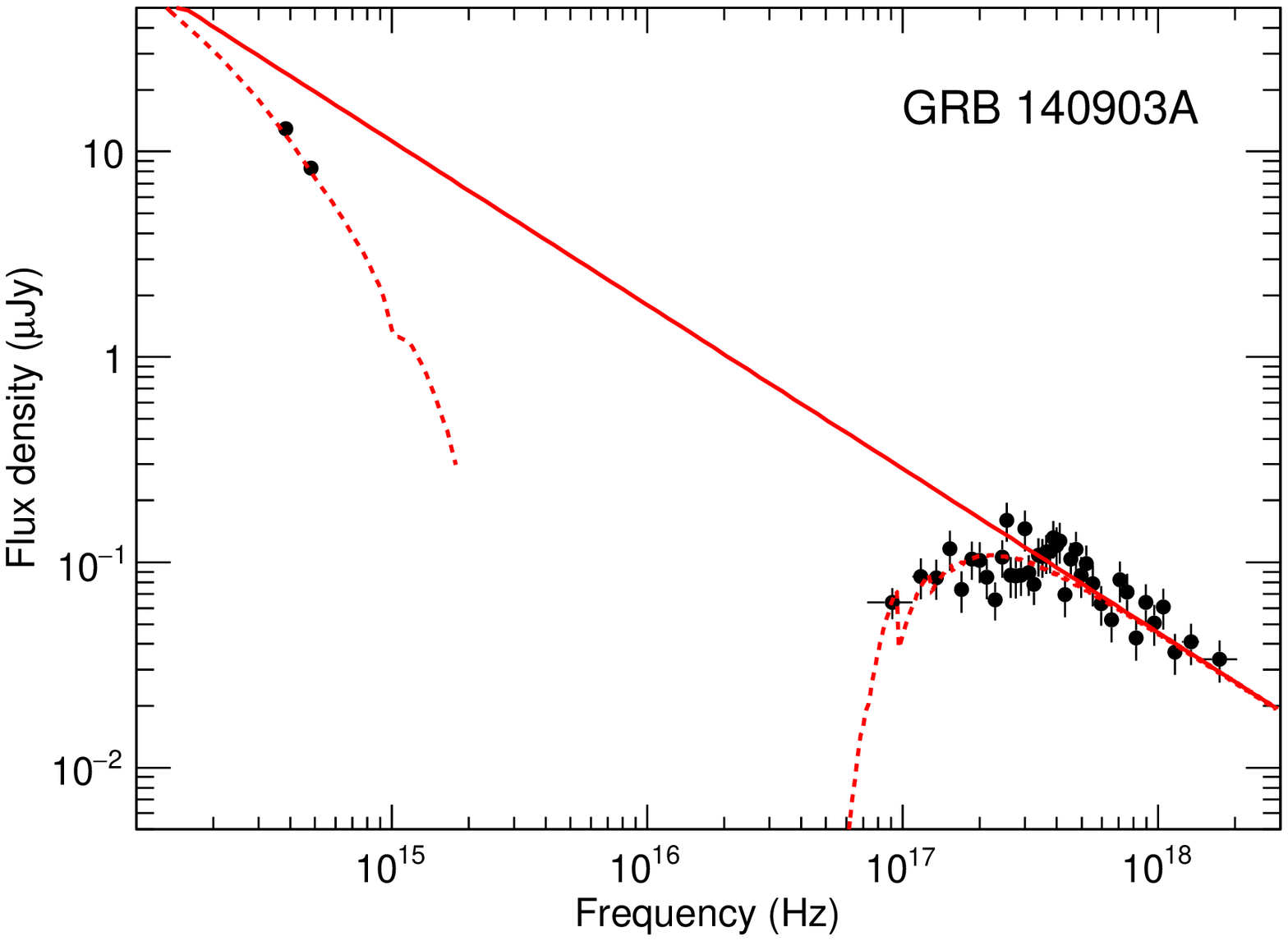}
    \end{center}
   \end{minipage}\\
   \begin{minipage}{0.33\hsize}
    \begin{center}
     \includegraphics[width=1.06\hsize]{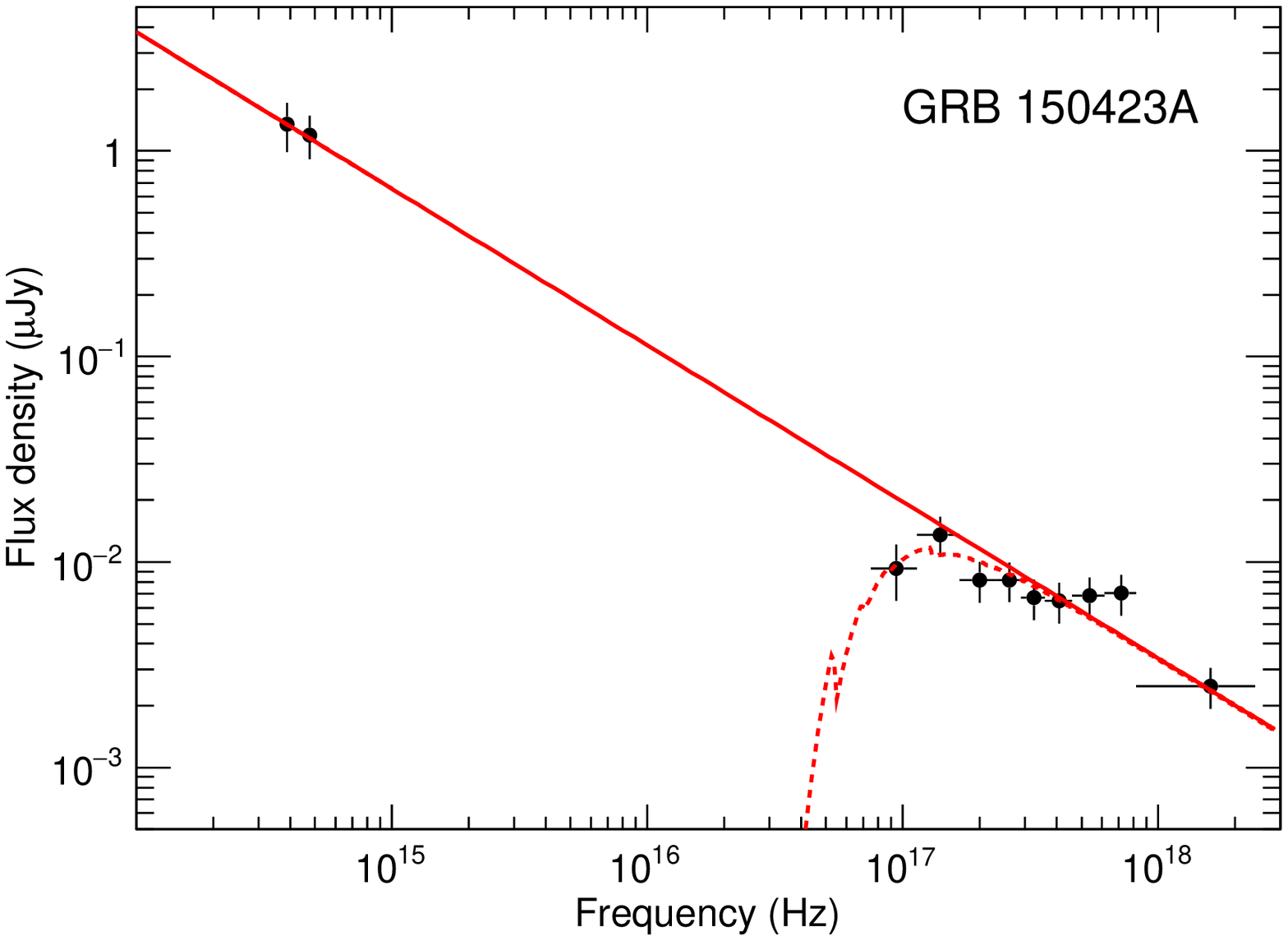}
    \end{center}
   \end{minipage}
   \begin{minipage}{0.33\hsize}
    \begin{center}
     \includegraphics[width=1.06\hsize]{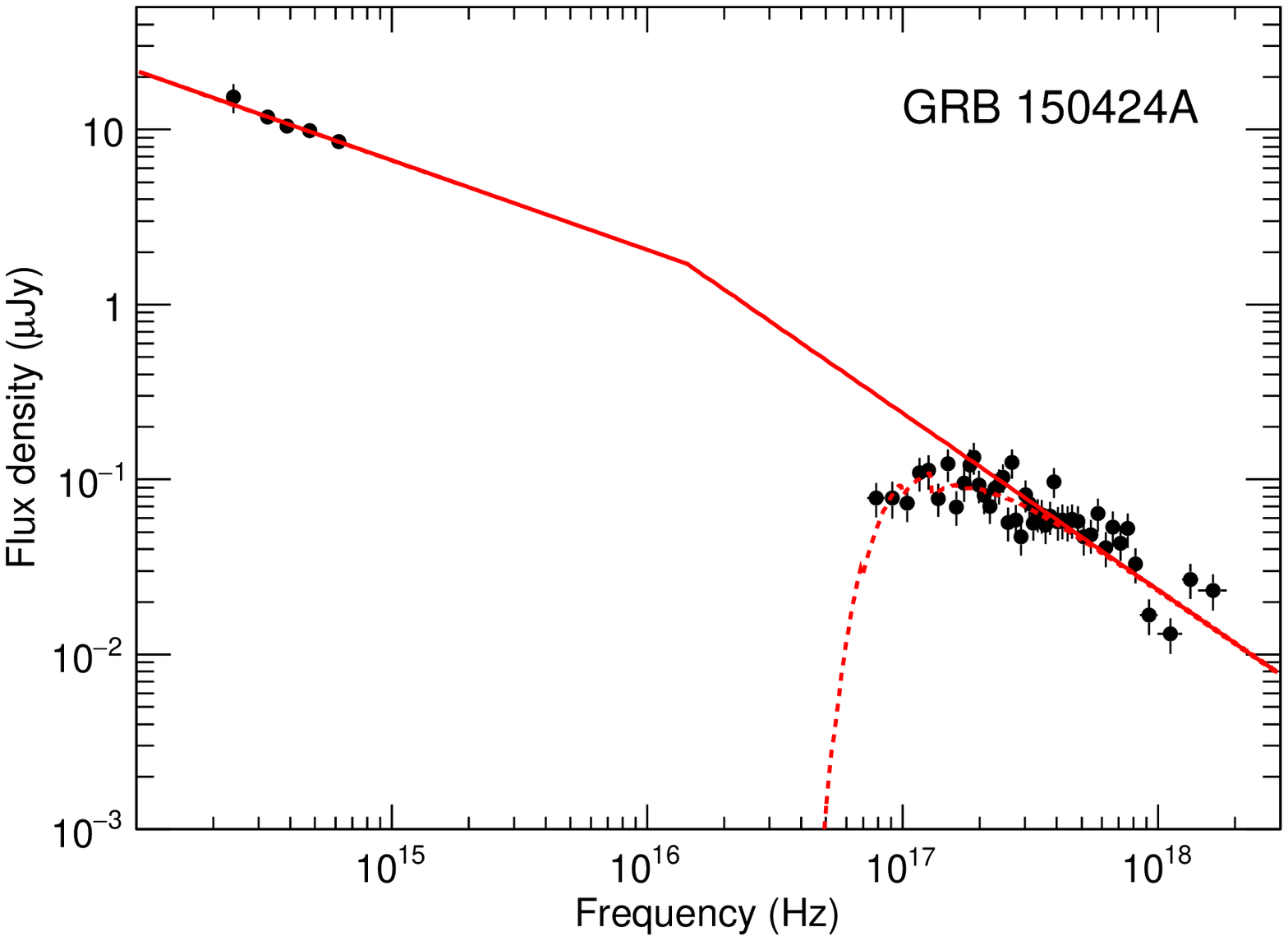}
    \end{center}
   \end{minipage}
   \begin{minipage}{0.33\hsize}
    \begin{center}
     \includegraphics[width=1.06\hsize]{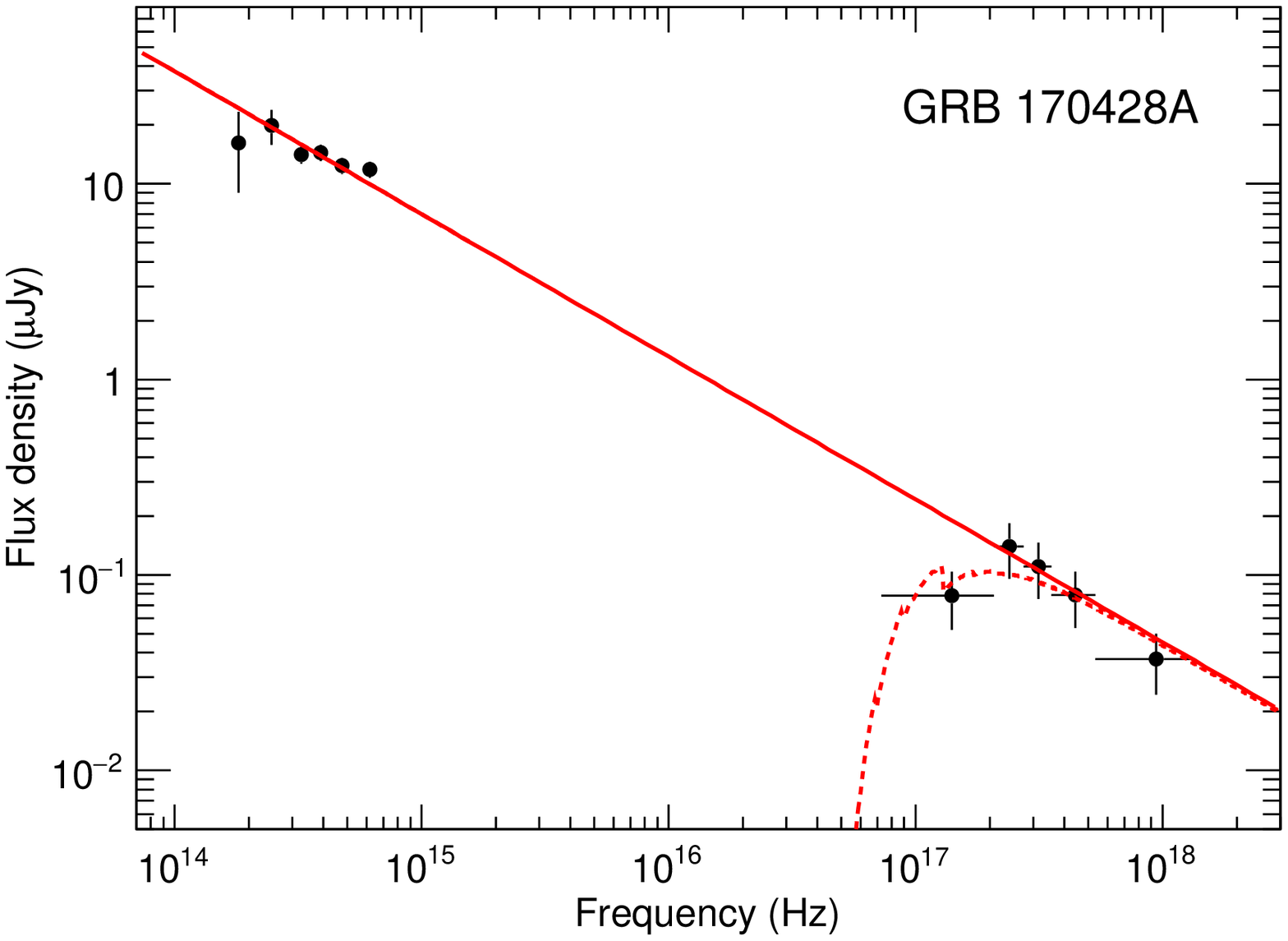}
    \end{center}
   \end{minipage}
  \end{tabular}
  \vspace{0.05in}
  \caption{The spectral energy distribution of 9~SGRBs. The optical/NIR
  data points are corrected for Galactic extinction, but the X-ray data
  points are not corrected for Galactic absorption. The solid lines show the best-fit unabsorbed spectral model corrected absorption and extinction. The dashed lines show the best-fit absorbed model including the Galactic and host-galactic absorption, and host-galactic extinction.\label{fig:spec}}
 \end{center}
\end{figure*}

The broadband SEDs with best-fit models are shown in Figure
\ref{fig:spec} and the results of our spectral analyses are summarized
in Table \ref{tab:result} (see also Table \ref{tab:all}). For the SEDs
of two SGRBs (GRB~130603B, 150424A), the broken power-law models have
better fitting results rather than the single power-law model. These are consistent
with the previous studies \citep{Postigo2014,Knust2017}.

\begin{figure}[h]
  \begin{center}
   \includegraphics[scale=0.45]{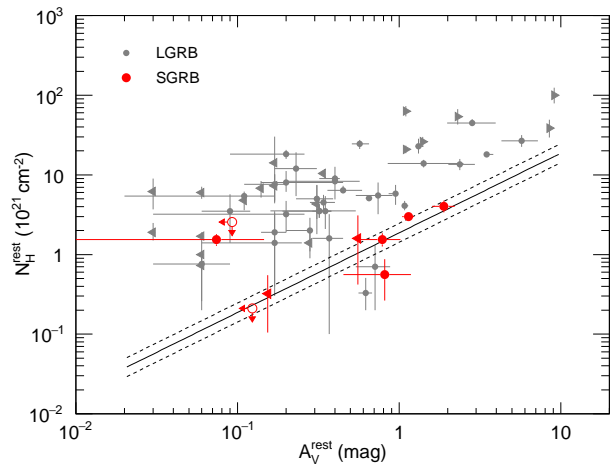}
   \caption{Rest-frame column density versus rest-frame extinction. The red
   and gray points are our result in this paper and the ones of LGRBs by
   \citet{Covino2013}. The triangles are the upper limits at $90\%$
   confidence level. The solid and dashed lines show the typical gas-to-dust ratio
   for the Milky Way and the corresponding $1\sigma$ uncertainty
   \citep{Welty2012}.\label{fig:sct}}
  \end{center}
 \end{figure}

Figure \ref{fig:sct} shows a scatter plot between $N_{\rm H}^{\rm rest}$
and $A_{\rm V}^{\rm rest}$ of SGRBs (this work) and LGRBs
\citep{Covino2013}, and the typical gas-to-dust ratio of the MW,
$N_{\rm H}/A_{\rm V}=1.9\times10^{21}$~cm$^{-2}$~mag$^{-1}$
\citep{Welty2012}. As shown in Figure \ref{fig:sct}, we found that the
$N_{\rm H}^{\rm rest}/A_{\rm V}^{\rm rest}$ ratio in the rest frame of
SGRBs is systematically smaller than the one of LGRBs, and is roughly
consistent with the gas-to-dust ratio in the MW.

\section{Discussion}
\begin{figure}[h]
\begin{center}
\includegraphics[scale=0.45]{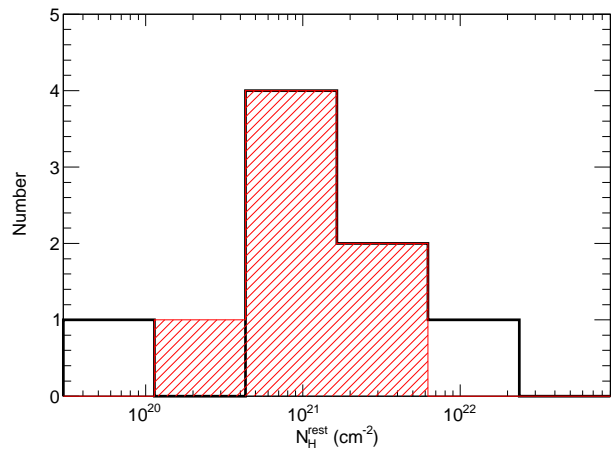}
\caption{Histgrams of the equivarent hydrogen column density in the GRB
 rest frame. The black and red lines show the distribution of our 9~SGRBs
 and additional 20~SGRBs, respectively.\label{fig:hist}}
\end{center}
\end{figure}

In order to investigate the selection effect on $N_{\rm H}^{\rm rest}$,
we analyzed X-ray afterglow spectra of all 20~SGRBs (not including our
9~samples) with known redshift observed by {\it Swift}/XRT before the
end of 2017, which did not have any near simultaneous
optical/NIR data. We performed the spectral analysis for each
time-averaged spectrum consists of observation data in PC mode. The
sample and the fitting result are listed in Table
\ref{tab:GRBlist2}. Figure \ref{fig:hist} shows the histograms of the
best fit value of $N_{\rm H}^{\rm rest}$ for our
initial 9~samples and additional 20~samples. We created
the cumulative distribution of best fit $N_{\rm H}^{\rm rest}$ and
applied the Kolmogorov-Smirnov test to it. Then, we found the null
hypothesis probability of 0.79 and our 9~samples show the same $N_{\rm
H}^{\rm rest}$ distribution of the other 20 SGRBs. Therefore we
concluded the $N_{\rm H}^{\rm rest}$ of our 9 SGRBs are not affected by
the selection bias, while we cannot give further argument on the
selection bias in $A_{\rm V}^{\rm rest}$ under the limited observation
data. Since \citet{Kruhler2011} reports the anti-correlation between the
$A_{\rm V}^{\rm rest}$ and the $N_{\rm H}^{\rm rest}/A_{\rm V}^{\rm
rest}$ ratio for LGRBs, the selection bias in $A_{\rm V}^{\rm rest}$ should be
discussed in detail for future observation data of SGRBs.

In our 9~SGRB samples, the measured gas-to-dust ratio
of SGRBs is roughly close to the one of the MW. Our result means that a major
contribution of both extinction in optical/NIR band and absorption in
X-ray band originates from the ISM in the host galaxy of SGRB. In other
words, most of SGRBs are likely to occur in not star-forming regions but
typical ISM environments of galaxies such as the MW. This result on the
environment is consistent with the scenario that the coalescence of the
compact binaries are the origin of because the system must move away
from the location of their birth by natal kicks until its merging
\citep[e.g.,][]{Narayan1992,Bloom1999,Fryer1999,Belczynski2006}.

$N_{\rm H}^{\rm rest}$ will show the amount of the intervening ISM within
the host galaxy. In our results, we found approximately half of SGRB
samples show $N_{\rm H}^{\rm rest}$ to be consistent with zero while we
obtained only marginal upper limit on them. These SGRBs are considered
to occur in outskirt or outside of the host galaxies in which there are
almost no X-ray absorption (and dust extinction) by the ISM. Moreover,
while GRB~170817A with GW~170817, whose origin is the binary neutron
star merger \citep{Abbott2017a,Goldstein2017,Savchenko2017}, which
occurred at only 1~$r_{e}$ from the center of the host galaxy. Here
$r_{e}$ is given by a S\'ersic model \citep{Ciotti1999}. However the
X-ray absorption and optical extinction in the host galaxy are not
significantly detected \citep{Levan2017,
Pooley2018}. This event might occur at the location
apart from the host galaxy toward the observer's side. The $N_{\rm
H}^{\rm rest}$ value might be an indicator of the offset along the line
of sight.

\section*{Acknowledgments}
We gratefully thank the anonymous referee for quick responses and helpful
comments, and we also acknowledge the quick and kind responses of the
editors. We also thank Yuu Niino for useful discussions. This work made
use of data supplied by the UK Swift Science Data Centre at the
University of Leicester, and is supported by JSPS KAKENHI Grant Number
JP17J00905 (KY), JP16H06342 (DY), JP18J13042 (YK), MEXT KAKENHI Grant
Number JP18H04580 (DY), JP17H06362 (MA), and Sakigake 2018 Project of
Kanazawa University (DY). MA acknowledges the support from the JSPS
Leading Initiative for Excellent Young Researchers program.

\newpage
 \begin{table}[h]
  \caption{Samples of SGRBs.}
  \label{tab:GRBlist}
  \begin{tabular}{lcccc}\hline
   GRB & z & $N_{\rm H}^{\rm gal}$ & $A_{\rm V}^{\rm gal}$ & Epoch\\
   & & ($10^{20}$ cm$^{-2}$) & (mag) & (s)\\\hline
   050724 & 0.258 & 27.7 & 1.61 & 41783\\
   051221A & 0.5465 & 7.52 & 0.18 & 184701\\
   070724A & 0.457 & 1.21 & 0.04 & 10872\\
   090510 & 0.903 & 1.77 & 0.05 & 28267\\
   130603B & 0.3564 & 2.1 & 0.06 & 52714\\
   140903A & 0.351 & 3.26 & 0.09 & 47117\\
   150423A & 1.394 & 1.77 & 0.08& 15300\\
   150424A & 0.3 & 6.02 & 0.16 & 57903\\
   170428A & 0.454 & 6.95 & 0.16 & 3660\\\hline
  \end{tabular}
 \end{table}

 \begin{table}[h]
  \caption{Results of spectral analysis.}
  \label{tab:result}
  \begin{tabular}{lcccccc}\hline
   GRB & $N_{\rm H}^{\rm rest}$ & $A_{\rm V}^{\rm rest}$ & $\beta_{\rm
   X}$ & $E_{\rm bk}$ & $\chi^{2}$ (dof) & Null hypothesis\\
   & ($10^{21}$ cm$^{-2}$) & (mag) & & (eV) & & probability\\\hline
   050724 & $<0.21$ & $<0.12$ & $-0.74^{+0.01}_{-0.01}$ & -- & 40 (31) & 0.121\\
   051221A & $0.56^{+0.31}_{-0.29}$ & $0.81^{+0.37}_{-0.36}$ &
	       $-0.83^{+0.06}_{-0.06}$ & -- & 44 (46) & 0.544\\
   070724A & $4.03^{+0.73}_{-0.63}$ & $1.89^{+0.31}_{-0.30}$ &
	       $-0.77^{+0.02}_{-0.02}$ & -- & 23 (19) & 0.226\\
   090510 & $1.53^{+0.28}_{-0.26}$ & $0.07^{+0.07}_{-0.07}$ &
	       $-0.84^{+0.02}_{-0.02}$ & -- & 107 (85) & 0.051\\
   130603B & $2.99^{+0.30}_{-0.36}$ & $1.14^{+0.10}_{-0.10}$ &
	       $-0.98^{+0.08}_{-0.07}$ & $8^{+19}_{-6}$ & 48 (49) & 0.498\\
   140903A & $1.53^{+0.31}_{-0.28}$ & $0.79^{+0.23}_{-0.24}$ &
	       $-0.80^{+0.03}_{-0.03}$ & -- & 49 (39) & 0.128\\
   150423A & $1.59^{+1.50}_{-1.17}$ & $<0.55$ & $-0.76^{+0.03}_{-0.03}$
	       & -- & 6 (7) & 0.536\\
   150424A & $0.32^{+0.23}_{-0.22}$ & $<0.15$ & $-1.01^{+0.07}_{-0.07}$
	       & 59$^{+82}_{-34}$ & 66 (46) & 0.027\\ 
   170428A & $<2.55$ & $<0.09$ & $-0.73^{+0.03}_{-0.02}$ & -- & 8 (7) & 0.344\\\hline
  \end{tabular}
 \end{table}

\begin{table}[h]
 \caption{Samples of additional 20~SGRBs.}
 \label{tab:GRBlist2}
 \begin{tabular}{lccc}\hline
  GRB & z & $N_{\rm H}^{\rm gal}$ & $N_{\rm H}^{\rm rest}$\\
  & & ($10^{20}$ cm$^{-2}$) & ($10^{21}$ cm$^{-2}$)\\\hline
  060614 & 0.125 & 2.09 & $0.11^{-0.01}_{-0.01}$\\
  060801 & 1.131 & 1.45 & $<1.4$\\
  061006 & 0.4377 & 25.1 & $<2.3$\\
  061201 & 0.111 & 6.8 & $<0.32$\\
  070714B & 0.923 & 9.82 & $0.87_{-0.57}^{+0.62}$\\
  070809 & 0.2187 & 8.62 & $<1.1$\\
  071227 & 0.383 & 1.31 & $<3.0$\\
  080123 & 0.495 & 2.52 & $<1.7$\\
  080905 & 0.121 & 13.5 & $1.28_{-0.77}^{+0.91}$\\
  090426 & 2.609 & 1.58 & $<3.0$\\
  090530 & 1.266 & 1.84 & $2.20_{-0.76}^{+0.81}$\\
  100117A & 0.915 & 2.97 & $1.11_{-0.97}^{+1.13}$\\
  100625A & 0.453 & 2.23 & $<0.66$\\
  100816A & 0.804 & 5.70 & $1.24_{-0.57}^{+0.63}$\\
  101219A & 0.718 & 5.91 & $4.39_{-3.35}^{+3.69}$\\
  111117A & 2.211 & 4.12 & $17.5_{-8.1}^{+9.8}$\\
  160228A & 1.64 & 8.98 & $<11$\\
  160410A & 1.717 & 1.8 & $<11$\\
  160624A & 0.483 & 9.31 & $<17$\\
  160821B & 0.16 & 5.95 & $<0.53$\\\hline
 \end{tabular}
\end{table}

\newpage

\clearpage

\appendix
\setcounter{table}{0}
\renewcommand{\thetable}{A.\arabic{table}}
\section{Optical/NIR observation data and result of spectral analysis}
\begin{ThreePartTable}
 \begin{TableNotes}
  \footnotesize	
  \item [1] Time since the trigger time (sec).
  \item [2] If not specified, the flux is not corrected for extinctions of our galaxy or the host one in the direction of the GRB. 
  \item [3] (a) \cite{Berger2005}, (b) \cite{Malesani2007}, (c)
  \cite{Soderberg2006}, (d) \cite{Berger2009}, (e) \cite{Fong2015} (f)
  \cite{Guelbenzu2012}, (g) \cite{Postigo2014}, (h) \cite{Troja2016},
  (i) \cite{Varela2015}, (j) \cite{Littlejohns2015}, (k) \cite{Kann2015},  (l) \cite{Knust2017}, (m) \cite{Bolmer2017}, (n) \cite{Troja2017}
  \item [4] The fluxes are corrected for Galactic extinction in the direction of the GRB. 
 \end{TableNotes}
 \begin{longtable}{lcrcc}	
  \caption{Optical/NIR observation data of our sample}
  \label{tab:opt} \\
  \toprule
  GRB & Filter & $\delta t$\tnote{1} ~(sec) & Flux\tnote{2} ~($\mu$Jy) & Reference\tnote{3} \\ 
  \midrule
  \endfirsthead
  \caption{continued} \\
  \toprule 
  GRB & Filter & $\delta t$\tnote{1} ~(sec) & Flux\tnote{2} ~($\mu$Jy) & Reference\tnote{3} \\ 
  \midrule
  \endhead
  \bottomrule
  \endfoot
  \bottomrule
  \insertTableNotes  
  \endlastfoot
  
  \input{./Yoshida+2018_big_table_rev3}
 \end{longtable} 
\end{ThreePartTable}

\begin{ThreePartTable}
 \begin{TableNotes}
  \footnotesize	
  \item [*] Break energy are restricted by the lower limit we set.
 \end{TableNotes}
 \begin{longtable}{llcccccc}	
  \caption{Results of spectral analysis for all model fit}
  \label{tab:all} \\
  \toprule
  GRB & Model& $N_{\rm H}^{\rm rest}$ & $A_{\rm V}^{\rm rest}$ & $\beta_{\rm
  X}$ & $E_{\rm bk}$ & $\chi^{2}/(dof)$ & Null hypothesis\\
  & & $(10^{21}~cm^{-2})$ & (mag) & & (eV) & & probability\\
  \midrule
  \endfirsthead
  \caption{continued}\\
  \toprule 
  GRB & Model &$N_{\rm H}^{\rm rest}$ & $A_{\rm V}^{\rm rest}$ & $\beta_{\rm
  X}$ & $E_{\rm bk}$ & $\chi^{2}/(dof)$ & Null hypothesis\\
  & & $(10^{21}~cm^{-2})$ & (mag) & & (eV) & & probability\\
  \midrule
  \endhead
  \bottomrule
  \endfoot
  \bottomrule
  \insertTableNotes  
  \endlastfoot

  050724 & MW/po & $<0.21$ & $<0.12$ & $-0.74^{+0.01}_{-0.01}$ & -- & 40 (31) & 0.121\\ 
  & LMC/po & $<0.21$ & $<0.19$ & $-0.74^{+0.01}_{-0.01}$ & -- & 40 (31) & 0.121\\ 
  & SMC/po & $<0.21$ & $<0.21$ & $-0.74^{+0.01}_{-0.01}$ & -- & 40 (31) & 0.121\\ 
  & MW/bknpo & $<0.39$ & $0.52^{+0.15}_{-0.15}$ & $-0.87^{+0.07}_{-0.06}$ & $5^{+12}_{-3}$ & 37 (30) & 0.183\\ 
  & LMC/bknpo & $<0.39$ & $0.51^{+0.14}_{-0.15}$ & $-0.87^{+0.07}_{-0.06}$ & $5^{+12}_{-3}$ & 37 (30) & 0.182\\ 
  & SMC/bknpo & $<0.39$ & $0.51^{+0.15}_{-0.15}$ & $-0.87^{+0.07}_{-0.06}$ & $5^{+12}_{-3}$ & 37 (30) & 0.180\\ 
  051221A & MW/po & $0.56^{+0.31}_{-0.29}$ & $0.81^{+0.37}_{-0.36}$ & $-0.83^{+0.06}_{-0.06}$ & -- & 44 (46) & 0.544\\ 
  & LMC/po & $0.55^{+0.31}_{-0.29}$ & $0.78^{+0.35}_{-0.35}$ & $-0.83^{+0.06}_{-0.06}$ & -- & 44 (46) & 0.540\\ 
  & SMC/po & $0.51^{+0.30}_{-0.29}$ & $0.72^{+0.34}_{-0.34}$ & $-0.82^{+0.05}_{-0.05}$ & -- & 45 (46) & 0.522\\ 
  & MW/bknpo & $1.00^{+0.40}_{-0.37}$ & $<1.02$ & $-0.95^{+0.08}_{-0.08}$ & $68^{+153}_{-46}$ & 38 (45) & 0.743\\ 
  & LMC/bknpo & $0.99^{+0.40}_{-0.23}$ & $<0.98$ & $-0.95^{+0.09}_{-0.08}$ & $65^{+155}_{-53}$ & 38 (45) & 0.743\\ 
  & SMC/bknpo & $0.98^{+0.41}_{-0.23}$ & $<0.88$ & $-0.94^{+0.09}_{-0.08}$ & $64^{+157}_{-50}$ & 38 (45) & 0.743\\ 
  070724A & MW/po & $4.03^{+0.73}_{-0.63}$ & $1.89^{+0.31}_{-0.30}$ & $-0.77^{+0.02}_{-0.02}$ & -- & 23 (19) & 0.226\\
  & LMC/po & $4.02^{+0.73}_{-0.63}$ & $1.85^{+0.31}_{-0.29}$ & $-0.77^{+0.02}_{-0.02}$ & -- & 23 (19) & 0.227\\ 
  & SMC/po & $4.00^{+0.73}_{-0.63}$ & $1.92^{+0.33}_{-0.31}$ & $-0.77^{+0.02}_{-0.02}$ & -- & 23 (19) & 0.229\\ 
  & MW/bknpo & $4.54^{+1.20}_{-0.72}$ & $2.55^{+0.33}_{-0.36}$ & $-0.85^{+0.13}_{-0.05}$ & $2^{+14}_{-2}$ & 23 (18) & 0.199\\ 
  & LMC/bknpo & $4.54^{+1.21}_{-0.71}$ & $2.49^{+0.33}_{-0.36}$ & $-0.85^{+0.13}_{-0.05}$ & $2^{+15}_{-2}$ & 23 (18) & 0.199\\ 
  & SMC/bknpo & $4.51^{+1.26}_{-0.70}$ & $2.58^{+0.33}_{-0.39}$ & $-0.84^{+0.14}_{-0.08}$ & $2^{+17}_{-2}$ & 23 (18) & 0.196\\ 
  090510 & MW/po & $1.53^{+0.28}_{-0.26}$ & $0.07^{+0.07}_{-0.07}$ & $-0.84^{+0.02}_{-0.02}$ & -- & 107 (85) & 0.051\\ 
  & LMC/po & $1.53^{+0.28}_{-0.26}$ & $<0.19$ & $-0.84^{+0.02}_{-0.02}$ & -- & 108 (85) & 0.049\\ 
  & SMC/po & $1.53^{+0.28}_{-0.26}$ & $<0.18$ & $-0.84^{+0.02}_{-0.02}$ & -- & 108 (85) & 0.050\\ 
  & MW/bknpo & $1.53^{+0.28}_{-0.26}$ & $0.07^{+0.07}_{-0.07}$ & $-0.84^{+0.02}_{-0.02}$ & \footnotemark[$*$] & 107 (84) & 0.043\\ 
  & LMC/bknpo & $1.53^{+0.28}_{-0.26}$ & $<0.19$ & $-0.84^{+0.02}_{-0.02}$ & \footnotemark[$*$] & 108 (84) & 0.042\\ 
  & SMC/bknpo & $1.53^{+0.28}_{-0.26}$ & $<0.18$ & $-0.84^{+0.02}_{-0.02}$ & \footnotemark[$*$] & 108 (84) & 0.043\\ 
  130603B & MW/po & $2.43^{+0.24}_{-0.22}$ & $0.79^{+0.05}_{-0.05}$ & $-0.83^{+0.01}_{-0.01}$ & -- & 58 (50) & 0.206\\ 
  & LMC/po & $2.42^{+0.24}_{-0.22}$ & $0.76^{+0.05}_{-0.05}$ & $-0.82^{+0.01}_{-0.01}$ & -- & 57 (50) & 0.218\\ 
  & SMC/po & $2.39^{+0.24}_{-0.22}$ & $0.72^{+0.05}_{-0.05}$ & $-0.82^{+0.01}_{-0.01}$ & -- & 56 (50) & 0.253\\ 
  & MW/bknpo & $2.99^{+0.30}_{-0.36}$ & $1.14^{+0.10}_{-0.10}$ & $-0.98^{+0.08}_{-0.07}$ & $8^{+19}_{-6}$ & 48 (49) & 0.498\\ 
  & LMC/bknpo & $3.01^{+0.38}_{-0.36}$ & $1.09^{+0.09}_{-0.09}$ & $-0.98^{+0.08}_{-0.07}$ & $8^{+21}_{-6}$ & 48 (49) & 0.505\\ 
  & SMC/bknpo & $3.10^{+0.42}_{-0.20}$ & $0.99^{+0.09}_{-0.09}$ & $-1.00^{+0.08}_{-0.07}$ & $15^{+38}_{-11}$ & 49 (49) & 0.477\\ 
  140903A & MW/po & $1.53^{+0.31}_{-0.28}$ & $0.79^{+0.23}_{-0.24}$ & $-0.80^{+0.03}_{-0.03}$ & -- & 49 (39) & 0.128\\
  & LMC/po & $1.53^{+0.31}_{-0.28}$ & $0.76^{+0.22}_{-0.23}$ & $-0.80^{+0.03}_{-0.03}$ & -- & 49 (39) & 0.130\\ 
  & SMC/po & $1.51^{+0.30}_{-0.27}$ & $0.74^{+0.21}_{-0.22}$ & $-0.79^{+0.03}_{-0.03}$ & -- & 49 (39) & 0.139\\ 
  & MW/bknpo & $1.53^{+0.31}_{-0.28}$ & $0.79^{+0.23}_{-0.24}$ & $-0.80^{+0.03}_{-0.03}$ & \footnotemark[$*$] & 49 (38) & 0.106\\ 
  & LMC/bknpo & $1.53^{+0.31}_{-0.28}$ & $0.76^{+0.22}_{-0.23}$ & $-0.80^{+0.03}_{-0.03}$ & \footnotemark[$*$] & 49 (38) & 0.108\\ 
  & SMC/bknpo & $1.51^{+0.30}_{-0.27}$ & $0.74^{+0.21}_{-0.21}$ & $-0.79^{+0.03}_{-0.03}$ & \footnotemark[$*$] & 49 (38) & 0.116\\ 
  150423A & MW/po & $1.59^{+1.50}_{-1.17}$ & $<0.55$ & $-0.76^{+0.03}_{-0.03}$ & -- & 6 (7) & 0.536\\
  & LMC/po & $1.59^{+1.50}_{-1.17}$ & $<0.57$ & $-0.76^{+0.03}_{-0.03}$ & -- & 6 (7) & 0.536\\ 
  & SMC/po & $1.59^{+1.50}_{-1.17}$ & $<0.56$ & $-0.76^{+0.03}_{-0.03}$ & -- & 6 (7) & 0.536\\ 
  & MW/bknpo & $1.59^{+1.50}_{-1.17}$ & $<0.55$ & $-0.76^{+0.03}_{-0.03}$ & \footnotemark[$*$] & 6 (6) & 0.419\\ 
  & LMC/bknpo & $1.59^{+1.50}_{-1.17}$ & $<0.57$ & $-0.76^{-0.50}_{-0.50}$ & \footnotemark[$*$] & 6 (6) & 0.419\\ 
  & SMC/bknpo & $1.59^{+1.50}_{-1.17}$ & $<0.56$ & $-0.76^{+0.03}_{-0.03}$ & \footnotemark[$*$] & 6 (6) & 0.419\\ 
  150424A  & MW/po & $<0.09$ & $<0.03$ & $-0.76^{+0.01}_{-0.01}$ & -- & 89 (47) & 2.66e-02\\ 
  & LMC/po & $<0.09$ & $<0.03$ & $-0.76^{+0.01}_{-0.01}$ & -- & 89 (47) & 2.66e-02\\ 
  & SMC/po & $<0.09$ & $<0.03$ & $-0.76^{+0.01}_{-0.01}$ & -- & 89 (47) & 2.66e-02\\ 
  & MW/bknpo & $0.32^{+0.23}_{-0.22}$ & $<0.15$ & $-1.01^{+0.06}_{-0.06}$ & $59^{+82}_{-34}$ & 66 (46) & 0.027\\ 
  & LMC/bknpo & $0.32^{+0.23}_{-0.22}$ & $<0.16$ & $-1.01^{+0.06}_{-0.06}$ & $59^{+82}_{-34}$ & 66 (46) & 0.027\\ 
  & SMC/bknpo & $0.32^{+0.23}_{-0.22}$ & $<0.15$ & $-1.01^{+0.06}_{-0.06}$ & $59^{+82}_{-34}$ & 66 (46) & 0.027\\ 
  170428A & MW/po & $<2.55$ & $<0.09$ & $-0.73^{+0.03}_{-0.02}$ & -- & 8 (7) & 0.344\\ 
  & LMC/po & $<2.55$ & $<0.06$ & $-0.73^{+0.03}_{-0.02}$ & -- & 8 (7) & 0.344\\ 
  & SMC/po & $<2.55$ & $<0.06$ & $-0.73^{+0.03}_{-0.02}$ & -- & 8 (7) & 0.344\\ 
  & MW/bknpo & $<3.72$ & $<0.28$ & $-0.92^{+0.16}_{-0.17}$ & $26^{+238}_{-23}$ & 2 (6) & 0.870\\ 
  & LMC/bknpo & $<3.72$ & $<0.22$ & $-0.91^{+0.17}_{-0.15}$ & $20^{+247}_{-18}$ & 3 (6) & 0.869\\ 
  & SMC/bknpo & $<3.72$ & $<0.21$ & $-0.92^{+0.16}_{-0.17}$ & $26^{+238}_{-23}$ & 2 (6) & 0.870\\ 
 \end{longtable} 
\end{ThreePartTable}

\end{document}

%% file: Yoshida+2018_big_table_rev3.tex
050724 & $K$ & $41760$ & $38.7^{+1.4}_{-1.4}$ & (a) \\
 & $I$ & $42517$ & $8.2^{+0.2}_{-0.2}$ & (b) \\
 &  & $125420$ & $1.3^{+0.1}_{-0.1}$ & (b) \\
 &  & $298980$ & $0.15^{+0.05}_{-0.04}$ & (b) \\
 & $R$ & $41797$ & $5.7^{+0.2}_{-0.2}$ & (b) \\
 &  & $126160$ & $1.1^{+0.1}_{-0.1}$ & (b) \\
 & $V$ & $41070$ & $3.7^{+0.1}_{-0.1}$ & (b) \\
051221A & $z$ & $184697$ & $0.98^{+0.44}_{-0.30}$ & (c) \\
 & $i$ & $97986$ & $2.1^{+0.5}_{-0.4}$ & (c) \\
 &  & $183522$ & $0.74^{+0.28}_{-0.21}$ & (c) \\
 & $r$ & $11120$ & $14.6^{+1.1}_{-1.0}$ & (c) \\
 &  & $12277$ & $13.6^{+1.0}_{-1.0}$ & (c) \\
 &  & $97001$ & $2.2^{+0.2}_{-0.2}$ & (c) \\
 &  & $185890$ & $0.80^{+0.09}_{-0.08}$ & (c) \\
 &  & $272419$ & $0.82^{+0.24}_{-0.19}$ & (c) \\
 &  & $445116$ & $0.43^{+0.09}_{-0.08}$ & (c) \\
070724A\tnote{4} & $K$ & $10080$ & $9.3^{+1.5}_{-1.5}$ & (d) \\
 &  & $13320$ & $8.9^{+1.5}_{-1.5}$ & (d) \\
 & $H$ & $12240$ & $7.8^{+0.4}_{-0.4}$ & (d, e) \\
 & $J$ & $11160$ & $3.4^{+0.3}_{-0.3}$ & (d, e) \\
 & $i$ & $8280$ & $1.1^{+0.1}_{-0.1}$ & (d) \\
090510 & $z$ & $22299$ & $7.4^{+5.0}_{-3.0}$ & (f) \\
 &  & $22401$ & $11.3^{+5.0}_{-3.5}$ & (f) \\
 &  & $22609$ & $9.9^{+4.4}_{-3.1}$ & (f) \\
 &  & $22743$ & $9.5^{+2.8}_{-2.2}$ & (f) \\
 &  & $23639$ & $5.3^{+1.8}_{-1.4}$ & (f) \\
 &  & $24093$ & $4.0^{+2.0}_{-1.4}$ & (f) \\
 &  & $24984$ & $3.6^{+1.4}_{-1.0}$ & (f) \\
 &  & $25889$ & $2.4^{+1.2}_{-0.8}$ & (f) \\
 &  & $26335$ & $6.0^{+1.4}_{-1.2}$ & (f) \\
 &  & $27234$ & $3.5^{+1.1}_{-0.8}$ & (f) \\
 &  & $28125$ & $2.6^{+1.0}_{-0.7}$ & (f) \\
 &  & $28569$ & $3.2^{+1.1}_{-0.8}$ & (f) \\
 &  & $29024$ & $4.5^{+1.1}_{-0.9}$ & (f) \\
 &  & $29475$ & $2.8^{+1.3}_{-0.9}$ & (f) \\
 &  & $30375$ & $2.3^{+1.0}_{-0.7}$ & (f) \\
 &  & $30831$ & $2.4^{+1.3}_{-0.8}$ & (f) \\
 &  & $31725$ & $3.4^{+0.8}_{-0.7}$ & (f) \\
 &  & $32628$ & $1.9^{+1.0}_{-0.7}$ & (f) \\
 &  & $33077$ & $2.3^{+0.7}_{-0.5}$ & (f) \\
 &  & $35270$ & $1.5^{+0.7}_{-0.5}$ & (f) \\
 &  & $35715$ & $2.1^{+0.9}_{-0.7}$ & (f) \\
 & $i$ & $22609$ & $6.6^{+2.9}_{-2.0}$ & (f) \\
 &  & $22931$ & $5.0^{+2.1}_{-1.5}$ & (f) \\
 &  & $23127$ & $6.5^{+2.5}_{-1.8}$ & (f) \\
 &  & $23313$ & $6.9^{+2.5}_{-1.8}$ & (f) \\
 &  & $23639$ & $4.2^{+1.4}_{-1.0}$ & (f) \\
 &  & $24093$ & $4.9^{+1.0}_{-0.8}$ & (f) \\
 &  & $24540$ & $2.8^{+1.4}_{-0.9}$ & (f) \\
 &  & $24984$ & $3.5^{+1.3}_{-0.9}$ & (f) \\
 &  & $25443$ & $2.6^{+1.1}_{-0.8}$ & (f) \\
 &  & $25889$ & $2.3^{+1.1}_{-0.7}$ & (f) \\
 &  & $26780$ & $2.0^{+0.6}_{-0.4}$ & (f) \\
 &  & $27234$ & $3.6^{+0.7}_{-0.6}$ & (f) \\
 &  & $27679$ & $3.4^{+0.6}_{-0.5}$ & (f) \\
 &  & $28125$ & $2.7^{+0.7}_{-0.6}$ & (f) \\
 &  & $28569$ & $2.9^{+0.8}_{-0.6}$ & (f) \\
 &  & $29024$ & $2.7^{+0.7}_{-0.6}$ & (f) \\
 &  & $29475$ & $2.4^{+0.7}_{-0.6}$ & (f) \\
 &  & $29922$ & $2.3^{+0.8}_{-0.6}$ & (f) \\
 &  & $30375$ & $2.2^{+0.7}_{-0.5}$ & (f) \\
 &  & $30831$ & $2.5^{+0.8}_{-0.6}$ & (f) \\
 &  & $31275$ & $1.9^{+0.4}_{-0.3}$ & (f) \\
 &  & $31725$ & $2.1^{+0.4}_{-0.4}$ & (f) \\
 &  & $32170$ & $1.3^{+0.6}_{-0.4}$ & (f) \\
 &  & $32628$ & $2.3^{+0.7}_{-0.5}$ & (f) \\
 &  & $33077$ & $1.8^{+0.6}_{-0.4}$ & (f) \\
 &  & $33524$ & $2.3^{+0.3}_{-0.3}$ & (f) \\
 &  & $34369$ & $1.1^{+0.7}_{-0.4}$ & (f) \\
 &  & $35270$ & $1.9^{+0.6}_{-0.4}$ & (f) \\
 &  & $35715$ & $1.0^{+0.7}_{-0.4}$ & (f) \\
 & $r$ & $22299$ & $5.7^{+2.4}_{-1.7}$ & (f) \\
 &  & $22503$ & $5.3^{+2.2}_{-1.6}$ & (f) \\
 &  & $22743$ & $4.4^{+1.6}_{-1.2}$ & (f) \\
 &  & $22931$ & $2.5^{+1.8}_{-1}$ & (f) \\
 &  & $23127$ & $2.9^{+1.8}_{-1.1}$ & (f) \\
 &  & $23313$ & $2.5^{+1.6}_{-1.0}$ & (f) \\
 &  & $24093$ & $2.2^{+0.7}_{-0.5}$ & (f) \\
 &  & $24540$ & $3.7^{+0.7}_{-0.6}$ & (f) \\
 &  & $24984$ & $2.6^{+0.9}_{-0.6}$ & (f) \\
 &  & $25443$ & $3.3^{+0.8}_{-0.6}$ & (f) \\
 &  & $25889$ & $2.8^{+0.7}_{-0.6}$ & (f) \\
 &  & $26335$ & $2.1^{+0.7}_{-0.5}$ & (f) \\
 &  & $26780$ & $3.0^{+0.7}_{-0.6}$ & (f) \\
 &  & $27234$ & $2.2^{+0.6}_{-0.4}$ & (f) \\
 &  & $27679$ & $1.8^{+0.4}_{-0.3}$ & (f) \\
 &  & $28125$ & $1.9^{+0.5}_{-0.4}$ & (f) \\
 &  & $28569$ & $2.1^{+0.5}_{-0.4}$ & (f) \\
 &  & $29024$ & $1.7^{+0.6}_{-0.4}$ & (f) \\
 &  & $29475$ & $2.0^{+0.5}_{-0.4}$ & (f) \\
 &  & $29922$ & $1.9^{+0.5}_{-0.4}$ & (f) \\
 &  & $30375$ & $2.1^{+0.4}_{-0.4}$ & (f) \\
 &  & $30831$ & $2.3^{+0.5}_{-0.4}$ & (f) \\
 &  & $31275$ & $1.9^{+0.5}_{-0.4}$ & (f) \\
 &  & $31725$ & $1.8^{+0.4}_{-0.3}$ & (f) \\
 &  & $32170$ & $1.5^{+0.5}_{-0.3}$ & (f) \\
 &  & $32628$ & $1.6^{+0.4}_{-0.3}$ & (f) \\
 &  & $33077$ & $1.2^{+0.4}_{-0.3}$ & (f) \\
 &  & $33524$ & $1.4^{+0.4}_{-0.3}$ & (f) \\
 &  & $34369$ & $1.4^{+0.4}_{-0.3}$ & (f) \\
 &  & $34815$ & $0.86^{+0.45}_{-0.30}$ & (f) \\
 &  & $35270$ & $1.2^{+0.4}_{-0.3}$ & (f) \\
 & $g$ & $23127$ & $2.6^{+1.7}_{-1.0}$ & (f) \\
 &  & $23639$ & $2.5^{+1.0}_{-0.7}$ & (f) \\
 &  & $24984$ & $2.6^{+0.9}_{-0.7}$ & (f) \\
 &  & $25889$ & $2.1^{+0.7}_{-0.5}$ & (f) \\
 &  & $26335$ & $1.5^{+0.7}_{-0.5}$ & (f) \\
 &  & $26780$ & $1.0^{+0.6}_{-0.4}$ & (f) \\
 &  & $27234$ & $1.4^{+0.6}_{-0.4}$ & (f) \\
 &  & $27679$ & $1.6^{+0.4}_{-0.3}$ & (f) \\
 &  & $28125$ & $1.3^{+0.5}_{-0.3}$ & (f) \\
 &  & $29024$ & $1.4^{+0.5}_{-0.4}$ & (f) \\
 &  & $29475$ & $1.5^{+0.5}_{-0.4}$ & (f) \\
 &  & $31275$ & $1.2^{+0.4}_{-0.3}$ & (f) \\
 &  & $31725$ & $1.1^{+0.4}_{-0.3}$ & (f) \\
 &  & $32170$ & $0.99^{+0.33}_{-0.25}$ & (f) \\
 &  & $32628$ & $1.1^{+0.4}_{-0.3}$ & (f) \\
 &  & $33077$ & $1.4^{+0.3}_{-0.3}$ & (f) \\
 &  & $34369$ & $0.60^{+0.41}_{-0.24}$ & (f) \\
 &  & $34815$ & $0.77^{+0.25}_{-0.19}$ & (f) \\
 &  & $35270$ & $1.0^{+0.3}_{-0.3}$ & (f) \\
 &  & $35715$ & $0.75^{+0.37}_{-0.25}$ & (f) \\
130603B & $K$ & $52099$ & $13.7^{+1.5}_{-1.3}$ & (g) \\
 & $J$ & $53050$ & $9.3^{+1.3}_{-1.1}$ & (g) \\
 & $z$ & $21946$ & $25.4^{+1.4}_{-1.4}$ & (g) \\
 &  & $51754$ & $6.1^{+0.2}_{-0.2}$ & (g) \\
 & $i$ & $23674$ & $16.4^{+0.9}_{-0.9}$ & (g) \\
 &  & $52445$ & $4.2^{+0.1}_{-0.1}$ & (g) \\
 & $r$ & $21082$ & $12.6^{+0.2}_{-0.2}$ & (g) \\
 &  & $25056$ & $11.0^{+0.2}_{-0.2}$ & (g) \\
 &  & $53136$ & $2.7^{+0.1}_{-0.1}$ & (g) \\
 &  & $138240$ & $0.21^{+0.07}_{-0.05}$ & (g) \\
 & $g$ & $25402$ & $6.3^{+0.4}_{-0.3}$ & (g) \\
 &  & $53827$ & $1.5^{+0.1}_{-0.1}$ & (g) \\
140903A & $i$ & $51840$ & $10.7^{+0.5}_{-0.5}$ & (h) \\
 &  & $140832$ & $2.3^{+0.3}_{-0.3}$ & (h) \\
 & $r$ & $44064$ & $8.6^{+0.7}_{-0.6}$ & (h) \\
 &  & $45792$ & $8.1^{+0.5}_{-0.4}$ & (h) \\
 150423A & $z$ & $240$ & $3.3^{+0.7}_{-0.6}$ & (i) \\
 & $i$ & $240$ & $2.8^{+0.6}_{-0.5}$ & (i) \\
 &  & $15300$ & $1.3^{+0.4}_{-0.3}$ & (j) \\
 & $R$ & $5655$ & $2.4^{+0.2}_{-0.2}$ & (k) \\
 &  & $9255$ & $1.6^{+0.2}_{-0.1}$ & (k) \\
 & $r$ & $240$ & $2.1^{+0.4}_{-0.4}$ & (i) \\
 &  & $15300$ & $1.1^{+0.3}_{-0.2}$ & (j) \\
 & $g$ & $240$ & $1.9^{+0.4}_{-0.3}$ & (i) \\
 150424A & $J$ & $57929$ & $14.7 ^{+2.8 }_{-2.4 }$ & (l) \\
 &  & $62670$ & $11.8^{+2.3 }_{-1.9}$ & (l) \\
 &  & $67399$ & $10.2^{+2.4 }_{-1.9}$ & (l) \\
 & $z$ & $57903$ & $11.1^{+0.6}_{-0.6}$ & (l) \\
 &  & $62645$ & $10.0^{+0.5}_{-0.5}$ & (l) \\
 &  & $67374$ & $9.0^{+0.5}_{-0.5}$ & (l) \\
 &  & $156355$ & $3.3^{+0.4}_{-0.3}$ & (l) \\
 & $i$ & $57903$ & $9.5^{+0.5}_{-0.4}$ & (l) \\
 &  & $62645$ & $9.6^{+0.4}_{-0.3}$ & (l) \\
 &  & $67374$ & $8.5^{+0.4}_{-0.4}$ & (l) \\
 &  & $156355$ & $2.2^{+0.3}_{-0.3}$ & (l) \\
 & $r$ & $57903$ & $8.7^{+0.2}_{-0.2}$ & (l) \\
 &  & $62645$ & $8.3^{+0.2}_{-0.2}$ & (l) \\
 &  & $67277$ & $6.9^{+0.2}_{-0.2}$ & (l) \\
 &  & $156582$ & $2.1^{+0.1}_{-0.1}$ & (l) \\
 &  & $323218$ & $0.98^{+0.22}_{-0.18}$ & (l) \\
 & $g$ & $57903$ & $7.2^{+0.3}_{-0.3}$ & (l) \\
 &  & $62645$ & $6.3^{+0.2}_{-0.2}$ & (l) \\
 &  & $67277$ & $5.8^{+0.2}_{-0.2}$ & (l) \\
 &  & $156123$ & $1.7^{+0.2}_{-0.2}$ & (l) \\
170428 & $H$ & $3660$ & $15.8^{+7.1}_{-4.9}$ & (m) \\
 & $J$ & $3660$ & $19.1^{+3.9}_{-3.2}$ & (m) \\
 & $z$ & $3660$ & $13.2^{+1.3}_{-1.2}$ & (m) \\
 & $i$ & $3660$ & $13.2^{+1.3}_{-1.2}$ & (m) \\
 &  & $91692$ & $5.2^{+2.3}_{-1.6}$ & (n) \\
 & $r$ & $3660$ & $11.0^{+1.1}_{-1.0}$ & (m) \\
 & $g$ & $3660$ & $10.0^{+1.0}_{-0.9}$ & (m) \\